\documentclass[journal,twoside]{IEEEtran}
\newcommand{\qeda}{\hfill\ensuremath{\blacksquare}}
\newcommand{\f}{it follows that }

\newcommand{\m}[1]{\mbox{$#1$}}

\ifCLASSINFOpdf
\else
\fi
\usepackage{multirow}

\usepackage{url}
\usepackage{cite}
\usepackage{amsfonts}
\usepackage{mathrsfs}
\usepackage{csquotes}
\usepackage{bbm}
\usepackage{pifont}

\usepackage{amsfonts}
\usepackage{amsmath, amsthm}
\usepackage{tabularx}
\usepackage{listings}
\usepackage{graphicx}
\usepackage{float}
\usepackage{amssymb}
\usepackage{array}
\usepackage{fancyhdr}
\usepackage{cases}
 \usepackage{supertabular}
\usepackage{relsize}
\usepackage{psfrag}
\usepackage{color}
\usepackage{bbding}

\newcommand {\n} {for any sufficiently large $n$}

\newcommand{\junzhao}[1]{\ding{110}\ding{43}\textcolor{red}{Jun Zhao: #1}}

\newcommand {\C} {{\rm I\kern-5.5pt C}}

\newcommand{\bcap} {\hspace{2pt} \mathlarger{\cap}
\hspace{2pt}}

\newcommand{\bP}[1]{{\mathbb{P}}\left[{#1}\right]}

\DeclareRobustCommand{\bEaa}{\protect\textnormal{static-network}^{\protect\begin{subarray}\textnormal{\textnormal{disk model}}\\ \textnormal{unit square}\end{subarray}}_{\textnormal{unreliable links}}}

\def\centerhack#1{\hbox to 0pt{\hss\footnotesize #1\hss}}
\def\centerhackn#1{\hbox to 0pt{\hss #1\hss}}
\def\dchack#1{\vbox to 0pt{\vss{\hbox to 0pt{\hss#1\hss}}\vss}}

\usepackage{enumitem}
\setlist{leftmargin=17pt}
\newcommand{\h}{it holds that }

\newtheorem{lem}{Lemma}
\newtheorem{thm}{Theorem}

\newtheorem{rem}{Remark}

\newtheorem*{proposition1.1}{Proposition 1.1}
\newtheorem*{proposition1.2}{Proposition 1.2}
\newtheorem*{proposition1.3}{Proposition 1.3}
\newtheorem*{proposition2.1}{Proposition 2.1}
\newtheorem*{proposition2.2}{Proposition 2.2}
\begin{document}
%
\title{Probabilistic Key Predistribution in Mobile Networks Resilient to
Node-Capture Attacks}
%
%
%

\author{Jun~Zhao,~\IEEEmembership{Member,~IEEE}
\thanks{Manuscript received August 18, 2015; revised October 24, 2016; accepted
June 20, 2017. Date of publication June 29, 2017; date of current version
September 13, 2017. This work was supported in part by Nanyang \mbox{Technological}
University, in part by Arizona State University, in part by the U.S.
Department of Energy, in part by the CyLab, and in part by the Department
of Electrical and Computer Engineering, Carnegie Mellon University. \newline \indent
The author was with the Cybersecurity Laboratory, Carnegie Mellon
University, Pittsburgh, PA 15213 USA, and with Arizona State University,
Tempe, AZ 85281 USA. He is now with Nanyang Technological University, Singapore
639798 (e-mail: junzhao@alumni.cmu.edu). \newline \indent
Communicated by Y. Liang, Associate Editor for Shannon Theory. \newline \indent
Color versions of one or more of the figures in this paper are available
online at http://ieeexplore.ieee.org. \newline \indent
Digital Object Identifier 10.1109/TIT.2017.2721424}}

%
%

\markboth{IEEE Transactions on Information Theory}%
{IEEE Transactions on Information Theory}
%



\maketitle


\begin{abstract}

We present a comprehensive analysis on
connectivity and resilience of secure sensor networks under the widely studied
$q$-composite key predistribution scheme.
 For network
connectivity which ensures that any two sensors can find a path in between for secure communication, we derive the conditions to guarantee connectivity in
consideration of (i) node-capture attacks, where the adversary may capture a set of sensors and compromise keys in their memory; (ii)  sensor mobility, meaning that sensors can move
around so that the network topology may change over time; (iii) physical transmission constraints, under
which two sensors have to be within each other's transmission range
for communication; (iv) the boundary effect of network fields; and
(v) link unreliability, meaning that links are allowed to be
unreliable. In contrast, many prior connectivity analyses of secure sensor networks often ignore the above issues. For resilience, although limited studies have presented formal analysis, it is often assumed that the adversary captures a random set of sensors, whereas this paper allows the adversary to capture an arbitrary set of sensors. We present     conditions
to ensure unassailability and unsplittability in secure sensor networks under the
$q$-composite scheme. Unassailability ensures that an adversary capturing any set consisting of a negligible fraction of sensors can
compromise only a negligible fraction of communication links although the adversary may compromise communications between
non-captured nodes which happen to use keys that are shared by
captured nodes. Unsplittability means that when a negligible fraction of sensors are captured, almost all of the remaining nodes are still securely connected. Based on the results of connectivity, unassailability and unsplittability, we provide useful guidelines for the
design of secure sensor networks.

\end{abstract}


%
\begin{IEEEkeywords}
Security, key predistribution,  wireless sensor networks, connectivity, random graphs.
 \end{IEEEkeywords}

%
%
%

\section{Introduction} \label{introduction}

\subsection{Background and Motivation}

Random key predistribution schemes have been widely recognized as
appropriate solutions to secure communications in
resource-constrained wireless sensor networks
\cite{adrian,virgil,Du2003CCS,yagan,yagan_onoff,pietro2004connectivity}.
The idea of randomly assigning cryptographic keys to sensors before
deployment was proposed in the seminal work of Eschenauer and
Gligor \cite{virgil}. The Eschenauer--Gligor (EG) scheme
\cite{virgil} works as follows. To secure a sensor network with $n$
nodes, in the key predistribution phase, the scheme uses a large
\emph{key pool} comprising   $P_n$ cryptographic keys to
select $K_n$ distinct keys \emph{uniformly at random} for each
sensor. These $K_n$ keys form the \emph{key ring} of a
sensor, and are inserted into the sensor's memory. After sensors are deployed,
two sensors can securely communicate over an existing link if
and only if their key rings share at least one key. Common
keys are found via neighbor discovery strategies \cite{virgil}. We let ${P}_n$ and
$K_n$ be functions of $n$ for generality, with the natural
condition $1 \leq K_n \leq P_n$.

Based on the EG scheme
\cite{virgil},
Chan \emph{et al.} \cite{adrian} propose the $q$-composite key
predistribution scheme as an extension. In such $q$-composite scheme, two sensors establish a secure link in between
 if and only if their key rings have at least $q$ key(s) in
common, where $1 \leq q \leq K_n$. Obviously, the
$q$-composite scheme with $q=1$ is the same as the EG scheme. The
$q$-composite scheme with $q\geq 2$ performs better than the EG scheme in
terms of resilience against small-scale node-capture attacks while
trading off heightened vulnerability in the presence of large-scale node-capture
attacks. The $q$-composite scheme has been
widely investigated in the literature over the last decade
 \cite{Rybarczyk,ToN17-topological,bloznelis2013,Perfectmatchings,7544484,Liu2003CCS,cha2017collaborative}.

For wireless sensor networks with probabilistic key predistribution, due to the analytical complexity,
most work in the literature on connectivity and resilience (explained below) considers only static networks with  few exceptions. In this
paper, we consider mobile networks in addition to static networks.
 An example application of probabilistic key predistribution to
mobile networks beyond static sensor networks is frequency hopping
\cite{MobiCom14}, which is a classic approach for transmitting
wireless signals by switching a carrier among different frequency
channels. In this application, each node uniformly and independently
selects $K_n$ secret seeds out of a secret pool consisting of $P_n$
secret seeds, and two nodes communicate with each other via
frequency hopping only if they share at least certain number of
secret seeds.

Although random key predistribution schemes enable secure communications in wireless sensor networks, they are often
not explicitly designed to defend against node-capture attacks, which
sensor networks deployed in hostile environments are often subject
to. The resilience of random key predistribution schemes to
node-capture attacks is often analyzed informally, and existing
results are often obtained under full visibility without addressing
physical transmission constraints, where the full visibility model
assumes that any two sensors have a direct communication link in
between. We present results on the resilience of the $q$-composite
scheme not only under full visibility, but also under physical
transmission constraints in which two nodes have to be within a
certain distance from each other to communicate.

%
%
%
%
%
%
%
%
%
%
%
%
%

A metric proposed by Chan \emph{et al.} \cite{adrian} to address
 the resilience to node-capture attacks is the probability (denoted by $p_{\textnormal{compromised}}$) that capturing
nodes enables an adversary to compromise communications between
non-captured nodes. Such probability $p_{\textnormal{compromised}}$
is non-zero in the EG scheme and its $q$-composite version because
non-captured nodes may happen to use keys that are also shared by
captured nodes \cite{4198829,DiPietroTissec}. Clearly, a random key
predistribution scheme is {perfectly resilient} to sensor-capture
attacks if $p_{\textnormal{compromised}}$ is alway zero. This implies that
only communications between a captured node and its direct neighbors
are compromised in a perfectly resilient scheme.

The EG scheme and its $q$-composite version can be easily extended
to become perfectly resilient in the Chan \emph{et al.}
\cite{adrian} sense \cite{JZNodeCapture,zhao2015resilience}. For instance, after the neighbor discovery phase
of the EG scheme ends, in case that two neighboring nodes identified by
$\textnormal{\emph{ID}}_a$ and $\textnormal{\emph{ID}}_b$ discover
that they share key $k_{ab}$, they can each use a cryptographic hash function $hash(\cdot)$ to compute a new shared key
$K_{ab} = hash(\textnormal{\emph{ID}}_a||
\textnormal{\emph{ID}}_b||k_{ab})$ for $a < b$, and
erase the old key $k_{ab}$.
 Since $K_{ab}$ is statistically unique up to the birthday bounds,
 the EG scheme becomes perfectly resilient. Similarly, the $q$-composite scheme can   become
perfectly resilient if the hash operation includes the uniquely ordered $q$ keys
 instead of the single key $k_{ab}$.

In mobile sensor networks employing the EG scheme or its
$q$-composite version, the key hashing process above is not
applicable, since sensors' neighbor sets change over time as sensors
move around. In static sensor networks with the key hashing process,
there is still a neighbor discovery phase where perfect resilience
cannot hold.
Therefore, it is of interest and significance to analyze the
resilience of $q$-composite scheme. 

In addition to analyzing the resilience of the $q$-composite scheme, we also investigate connectivity properties of secure sensor networks under  the $q$-composite scheme, when the adversary may capture a set of sensors.
Connectivity means that any two sensors can find a path in between for secure communication. Our connectivity analysis considers many different settings: the network can be static or mobile; we consider the case of no node capture as well as the case of node-capture attacks; limited visibility models considered include the disk model, the link unreliability model, and their combination; under the disk model, we consider both the case ignoring the boundary effect of the network fields as well as the case considering the boundary effect.

\subsection{Problems}  \label{motivation}

\subsubsection{Connectivity}  ~

Connectivity is a fundamental property that networks are often
designed to have.

\noindent \textbf{Definition of Connectivity.} \emph{A network is connected if
each node can find at least one path to reach another node.}

For secure sensor networks with the $q$-composite scheme, we  obtain
the conditions to have the networks connected; i.e., the conditions that any two sensors can find a path in between for secure communication.

\subsubsection{Unsplittability}  ~

Another metric to characterize resiliency is unsplittability
proposed by Pietro \emph{et al.} \cite{DiPietroTissec}. The motivation is that even if a negligible (i.e., $o(1)$)\footnote{All limits are taken with
$n\to\infty$, where $n$ is the number of nodes in a network. The
standard asymptotic notation $o(\cdot), O(\cdot), \omega(\cdot),
\Omega(\cdot), \Theta(\cdot), \sim$ are used. Given two positive
sequences $f_n$ and $g_n$, we have
\begin{enumerate}
  \item $f_n = o \left(g_n\right)$ means $\lim_{n \to
  \infty}\frac{f_n}{g_n}=0$.
  \item $f_n = O \left(g_n\right)$ means that there exist positive
  constants $c_1$ and $N_1$ such that $f_n \leq c_1 g_n$ for all $n \geq
  N_1$.
  \item $f_n = \omega \left(g_n\right)$ means $\lim_{n \to
  \infty}\frac{f_n}{g_n}=\infty$.
\item $f_n = \Omega \left(g_n\right)$ means that there exist positive
  constants $c_2$ and $N_2$ such that $f_n \geq c_2 g_n$ for all $n \geq
  N_2$. 
  \item $f_n = \Theta \left(g_n\right)$ means that there exist positive
  constants $c_3, c_4$ and $N_3$ such that
  $c_3 g_n \leq f_n \leq c_4 g_n$ for all $n \geq
  N_3$. 
  \item $f_n \sim g_n$ means that $\lim_{n \to
  \infty}\frac{f_n}{g_n}=1$; i.e., $f_n$
  and $g_n$ are asymptotically equivalent. %
\end{enumerate}
} fraction of nodes are compromised, almost all of the remaining nodes are still securely connected. Formally, unsplittability is defined as follows.

\noindent \textbf{Definition of Unsplittability.} \emph{A secure
sensor network is unsplittable if with high probability an adversary
that has captured an arbitrary set of $o(n)$ sensors cannot
partition the network into two chunks, which both have linear sizes
of sensors (i.e., $\Theta(n)$ sensors), and either are isolated from each other or only have
compromised communications in between (i.e., keys used for
communications between the two chucks are all compromised by the
adversary).}

From the definition above, for a unsplittable secure sensor network, even if $n_1 = o(n)$ nodes are compromised,   we still have a securely connected network consisting of $n-n_2$ nodes, where $n_2$ can also be written as $o(n)$ despite $n_2 \geq n_1$. The reason for $n_2 \geq n_1$ is that capturing
nodes enables an adversary to compromise communications between
non-captured nodes which happen to use keys that are also shared by
captured nodes.

For secure sensor networks with the $q$-composite scheme, we   explore
the conditions to have the networks unsplittable.

\subsubsection{Unassailability} ~

We borrow the notion of unassailability by Mei \emph{et al.} \cite{PanconesiUnassailable} and present the following definition. Briefly, a
secure sensor network is unassailable if
$p_{\textnormal{compromised}} $ is $o(1)$ after the adversary has
captured $o(n)$ sensors.

\noindent \textbf{Definition of Unassailability.} \emph{A secure sensor network is unassailable if
an
adversary that has captured an arbitrary set of $o(n)$
sensors can only compromise $o(1)$ fraction of communication links
in the rest of the network; in other words, an adversary has to capture a
constant fraction of sensors to compromise a constant fraction of
communication links.}

For secure sensor networks with the $q$-composite scheme, we derive
the conditions to have the networks unassailable.

Many prior studies have analyzed node-captured attacks, but they
often consider that the adversary \emph{randomly} capture sensors.
However, in practice, the adversary may tune its node-capture
strategy according to its goals. To analyze unassailability, we analyze \textit{arbitrary} node-capture attacks, in which the adversary can
capture an {arbitrary} set of sensors.
 The choice
of the captured nodes depends on the adversary's goal and also physical
limitations (e.g., some sensors may be easier to be captured than
others).

{

\begin{table}[t]
\centering
\begin{tabular}{|l|l|}
\hline
 \hspace{-4pt}\textbf{Symbols}\hspace{-4pt} & \hspace{-4pt}\textbf{Meanings}  \\ \hline
\hspace{-4pt}$P_n$ & \hspace{-4pt}the  size of the key pool  \\ \hline
\hspace{-4pt}$K_n$ & \hspace{-4pt}the number of keys assigned to each
sensor before deployment  \\ \hline
\hspace{-4pt}$q$ & \begin{tabular}[c]{@{}l@{}}\hspace{-4pt}the minimum number of keys that two sensors need to share\hspace{-4pt}\\ \hspace{-4pt}for secure communication over an existing link\end{tabular} \\ \hline
\hspace{-4pt}$r_n$ & \begin{tabular}[c]{@{}l@{}}\hspace{-4pt}the distance that two sensors have to be within each other\hspace{-4pt}\\ \hspace{-4pt}for communication under the disk model \end{tabular} \\ \hline
\hspace{-4pt}$t_n$ & \begin{tabular}[c]{@{}l@{}}\hspace{-4pt}the probability that each link is active when links are\hspace{-4pt}\\ \hspace{-4pt}allowed to be
 unreliable\end{tabular}   \\ \hline
\end{tabular}  \vspace{2pt}
  \caption{Symbols and their meanings.} \label{tabsym}
\end{table}
}

\subsection{Contributions}

Our work presents a rigorous and comprehensive analysis on
resilience and connectivity of secure sensor networks under the
$q$-composite key scheme. In particular, we investigate  conditions
to have unassailability and unsplittability. For network
connectivity, we derive the conditions to guarantee connectivity in
consideration of (i) node-capture attacks, where the adversary may capture a set of sensors and compromise keys in their memory; (ii)  sensor mobility, meaning that sensors can move
around so that the network topology may change over time; (iii) physical transmission constraints, under
which two sensors have to be within each other's transmission range
for communication; (iv) the boundary effect of network fields; and
(v) link unreliability, meaning that links are allowed to be
unreliable.

\textbf{Summary of Results.}
We summarize our results as follows. Note that all results are in the asymptotic sense.
\begin{itemize}
\item \textbf{Connectivity:}
We establish the following results for connectivity in secure sensor networks employing the $q$-composite scheme. Note that $r_n$ denotes the distance that two sensors have to be within each other for communication under the disk model, and $t_n$ denotes the probability that each link is active when links are allowed to be unreliable.\\
\textbf{Connectivity of static networks:}
\begin{itemize}
\item A static secure sensor network with the $q$-composite scheme under the disk model on the unit torus without the boundary effect is {connected}
if $\frac{1}{q!}  \cdot \frac{{K_n}^{2q}}{{P_n} ^{q}}  \cdot \pi
{r_n}^{2} \geq \frac{c\ln n}{n}$ for any constant $c>1$.
\item A static secure sensor network with the $q$-composite scheme and unreliable links under the disk model on the unit torus without the boundary effect is {connected}
if $\frac{1}{q!}  \cdot \frac{{K_n}^{2q}}{{P_n} ^{q}}  \cdot \pi
{r_n}^{2} \cdot t_n \geq \frac{c\ln n}{n}$ for any constant $c>1$.
\item A static secure sensor network with the $q$-composite scheme under the disk model on the unit square with the boundary effect is {connected}
if \mbox{$\frac{1}{q!}  \cdot \frac{{K_n}^{2q}}{{P_n} ^{q}}  \cdot \pi
{r_n}^{2}$}$ \geq \frac{c\ln n}{n}$ for any constant $c>c_n^{*} := \max $\mbox{$\{1 +   \ln \frac{P_n}{{K_n}^2} \big/ \ln n , $}$~4  \ln \frac{P_n}{{K_n}^2} \big/ \ln n  \} $.
\item A static secure sensor network with the $q$-composite scheme and unreliable links under the disk model on the unit square with the boundary effect is {connected}
if $\frac{1}{q!}  \cdot \frac{{K_n}^{2q}}{{P_n} ^{q}}  \cdot \pi
{r_n}^{2} \cdot t_n \geq    \frac{c\ln n}{n}$ for any constant $c> c_n^{\#} := \max \{1 + ( q \ln \frac{P_n}{{K_n}^2} + \ln \frac{1}{t_n} ) \big/ \ln n , ~4 ( q \ln \frac{P_n}{{K_n}^2} + \ln \frac{1}{t_n} ) \big/ \ln n \} $.
\end{itemize}
\textbf{Connectivity of mobile networks:}
\begin{itemize}
\item A mobile secure sensor network with the $q$-composite scheme under the disk model on the unit torus without the boundary effect and under the i.i.d. mobility model is connected for at least $
n^{c-1-\epsilon}$ consecutive time slots from the beginning for an arbitrary positive constant $\epsilon$, if $\frac{1}{q!}  \cdot \frac{{K_n}^{2q}}{{P_n} ^{q}}  \cdot \pi
{r_n}^{2} \geq \frac{c\ln n}{n}$ for any constant $c>1$.
\item A mobile secure sensor network with the $q$-composite scheme and unreliable links under the disk model on the unit torus without the boundary effect and under the i.i.d. mobility model is connected for at least $
n^{c-1-\epsilon}$ consecutive time slots from the beginning for an arbitrary positive constant $\epsilon$, if $\frac{1}{q!}  \cdot \frac{{K_n}^{2q}}{{P_n} ^{q}}  \cdot \pi
{r_n}^{2}  \cdot t_n \geq \frac{c\ln n}{n}$ for any constant $c>1$.
\item A mobile secure sensor network with the $q$-composite scheme under the disk model on the unit square with the boundary effect and under the i.i.d. mobility model is connected for at least $
n^{c-c_n^{*}-\epsilon}$ consecutive time slots from the beginning for an arbitrary positive constant $\epsilon$, if $\frac{1}{q!}  \cdot \frac{{K_n}^{2q}}{{P_n} ^{q}}  \cdot \pi
{r_n}^{2} \geq   \frac{c\ln n}{n}$ for any constant $c> c_n^{*}$ with $c_n^{*}$ defined above.
\item A mobile secure sensor network with the $q$-composite scheme and unreliable links under the disk model on the unit square with the boundary effect and under the i.i.d. mobility model is connected for at least $
n^{c-c_{\#}-\epsilon}$ consecutive time slots from the beginning for an arbitrary positive constant $\epsilon$, if $\frac{1}{q!}  \cdot \frac{{K_n}^{2q}}{{P_n} ^{q}}  \cdot \pi
{r_n}^{2}  \cdot t_n  \geq  \frac{c\ln n}{n}$ for any constant $c> c_n^{\#}$ with $c_n^{\#}$ defined above.
\end{itemize}
\textbf{Connectivity under node capture:} To obtain the
connectivity results for an $n$-size
network after the adversary has captured a random set of $m$ nodes
out of all $n$ nodes, we just replace $\ln n$ with $\ln (n-m)$,   $\frac{\ln
n}{n}$ with $\frac{\ln (n-m)}{n-m}$, and $n$ with $(n-m)$ in the above
connectivity results (note that we do not replace the ``$n$'' in $K_n$, $P_n$, $r_n$, $t_n$).
\item \textbf{Unassailability:}
\begin{itemize}
\item A secure sensor network with the $q$-composite scheme under any
communication model is \emph{unassailable} if ${P_n}/{K_n} =
\Omega(n)$.
\end{itemize}
\item \textbf{Unsplittability:}
\begin{itemize}
\item A secure sensor network with the $q$-composite scheme under full visibility is \emph{unsplittable}
if $\frac{1}{q!}  \cdot \frac{{K_n}^{2q}}{{P_n} ^{q}} \geq \frac{c\ln n}{n}$ for any constant $c>1$, $K_n = \omega( \ln n)$ and $K_n =
o\big(\min\{\sqrt{P_n}, \frac{P_n}{n}\}\big)$.
\item A static secure sensor network with the $q$-composite scheme under the disk model on some area $\mathcal{A}$ is \emph{splittable}
if $r_n = o(\sqrt{\mathcal{A}})$.
\end{itemize}
\end{itemize}

\subsection{Roadmap}

The rest of the paper is organized as follows. Section
\ref{sec:related-work} reviews related work. In Section
\ref{sysmodel}, we describe the system model. Section
\ref{sec:main:res} presents the main results.  We provide several useful lemmas in Section \ref{sec-preliminary-results}. Afterwards, we
establish the theorems in Section \ref{prove:thm}. We explain
simulation results in Section \ref{sec:expe}. Finally, Section
\ref{sec:Conclusion} concludes the paper.

\section{Related Work} \label{sec:related-work}

\textbf{Unassailability.}
Mei \emph{et al.} \cite{PanconesiUnassailable} introduce the notion of
unassailability. However, their work is different from ours in the following aspects: they address the EG scheme whereas we consider the more general $q$-composite scheme; they assume that the adversary knows the key rings of all nodes whereas we do not have such assumption; they enforce the strong assumption that whenever the adversary has compromised  one shared key between two sensors, the secure link between the two sensors is compromised even though that compromised key is not used for the secure link. For the EG scheme, Di Pietro \emph{et al.}
\cite{4198829,DiPietroTissec} have presented conditions to ensure that an adversary has to capture a constant fraction of sensors in order to
compromise a constant fraction of communication links. Similar results for the $q$-composite scheme have been obtained by Zhao \cite{JZNodeCapture,zhao2015resilience}. However, the difference between \cite{4198829,DiPietroTissec,JZNodeCapture,zhao2015resilience} and this paper is that \cite{4198829,DiPietroTissec,JZNodeCapture,zhao2015resilience} consider an adversary \emph{randomly} capturing sensors, while this paper investigates the more practical case of \textit{arbitrary} node-capture attacks, in which the adversary can
capture an {arbitrary} set of sensors. In addition to the EG scheme and the $q$-composite scheme, the
random pairwise key predistribution scheme of Chan \textit{et al.} \cite{adrian}
has also received much attention. The unassailability of the pairwise   scheme has recently been studied by Ya{\u{g}}an and Makowski \cite{YaganMakowskiToN}. The key predistribution schemes by Liu and Ning \cite{Liu2003CCS}
and Du \emph{et al.} \cite{Du2003CCS}
exhibit the threshold behavior in terms of resilience against node-capture attacks. Techniques to
improve the resilience against node capture have been studied in \cite{Alarifi:2006:DSN:1180345.1180359,Yang:2005:TRS:1062689.1062696,Vu:2010:SWS:1755688.1755703,Conti:2008:EPD:1352533.1352568,5717499,tague2008modeling}.

\textbf{Unsplittability.} For the EG scheme, its unsplittability has   been studied by \cite{DiPietroTissec}. Note that our unsplittability result is for the more general $q$-composite scheme. Yet, even when we apply our general result to the EG scheme, the obtained result will be stronger than that of \cite{DiPietroTissec}. Specifically, since our unsplittability condition for the $q$-composite scheme uses $\frac{1}{q!}
 \cdot \frac{{K_n}^{2q}}{{P_n} ^{q}}  \geq
\frac{c\ln n}{n}$ for a constant $c>1$, our unsplittability result applying to the EG scheme (i.e., the $q$-composite scheme with $q=1$) uses $ \frac{{K_n}^{2}}{{P_n}}  \geq
\frac{c\ln n}{n}$ for a constant $c>1$. In contrast, the unsplittability result of the EG scheme in  \cite{DiPietroTissec} requires $ \frac{{K_n}^{2}}{{P_n}}  \geq
\frac{c\ln n}{n}$ for a constant $c>17$. For the random pairwise key predistribution scheme of Chan \textit{et al.} \cite{adrian}, Ya{\u{g}}an and Makowski \cite{YaganMakowskiToN} have recently   studied  its unsplittability.

\textbf{Connectivity.} Although connectivity of secure sensor networks has been widely investigated in the literature, many prior studies has one or more than one of the following limitations: much research  \cite{pietro2004connectivity,yagan,adrian} ignores either real-world transmission constraints between sensors or mobility of sensors because analyzing   the key predistribution scheme, transmission constraints, and node mobility  together makes the analysis very challenging; several researches considering transmission constraints   obtain quite weak results \cite{ISIT_RKGRGG,Krzywdzi,Bloznelis201494,Perfectmatchings} or apply to limited settings \cite{JZNodeCapture} (e.g., the case ignoring the boundary effect of the network fields); the proof techniques   \cite{ANALCO} lack formality. Below we discuss some related work in detail.

For a secure sensor network with the $q$-composite scheme under the full visibility, its connectivity  has been studied in \cite{ANALCO,Perfectmatchings}, while its $k$-connectivity  has been considered in \cite{Bloznelis201494}, where $k$-connectivity ensures connectivity despite the failure of any $k-1$ nodes \cite{ZhaoYaganGligor,fu2017we}. However, the proof techniques in  \cite{ANALCO} lack formality (specifically \cite[Equation (6.93)]{ANALCO}) and are different from those in this paper.
The results of connectivity in \cite{Bloznelis201494} and $k$-connectivity in \cite{Perfectmatchings} address only the narrow and impractical range of $P_n =o\big(n^{\frac{1}{q}} (\ln n)^{-\frac{3}{5q}}\big)$, as explained below. In contrast, from condition set $\Lambda$ in (\ref{Lambda-set}), our Theorem \ref{Connect-full-visi} considers a more practical range of $P_n = \omega(n \ln n)$. We present more details below.

We explain that the  result of \cite{Perfectmatchings} for connectivity  is not applicable to practical sensor networks, whereas our result is applicable. The limitation of \cite{Perfectmatchings} is that both $K_n$ and $P_n$ are required to be quite small in \cite{Perfectmatchings} so they will not satisfy the resiliency requirement $\frac{P_n}{K_n} =
\Omega(n)$. More specifically, \cite[Equation (2)]{Perfectmatchings} enforces
  \begin{align}
&\left[(q+2)\binom{K_n}{q}^5 (\ln \ln n)^2\right]^{\frac{3K_n-q}{3(K_n-q)}}& \leq (\ln n)^{1-\gamma}\nonumber \\ & \text{for a constant }\gamma \in (0,1), \label{eqn-Perfectmatchings-1}
\end{align}
where the notation $s$ and $d$ in \cite{Perfectmatchings} correspond to $q$ and $K_n$ here.

In (\ref{eqn-Perfectmatchings-1}), the exponent $\frac{3K_n-q}{3(K_n-q)}$ is greater than $1$ given $K_n >q$, since $q$ is a constant and does not scale with $n$ ($q$ is often less than $10$), while $K_n$ controls the number of keys on each sensor and is at least a two-digit number for a network comprising thousands of sensors. In fact, if $K_n < q$, the $q$-composite scheme is meaningless because each sensor has just $K_n$ keys so it is impossible for two sensors to share $q$ keys if $K_n < q$. Furthermore, in (\ref{eqn-Perfectmatchings-1}), the base is greater than $1$ for all sufficiently large owing to $\lim_{n\to\infty} \ln \ln n = \infty$. Given the above, we can cancel out the exponent in (\ref{eqn-Perfectmatchings-1}); i.e., (\ref{eqn-Perfectmatchings-1}) implies
\begin{align}
&(q+2)\binom{K_n}{q}^5 (\ln \ln n)^2 \leq (\ln n)^{1-\gamma} \nonumber \\ & \text{for a constant }\gamma \in (0,1). \label{eqn-Perfectmatchings-2}
\end{align}
In view of  $\binom{K_n}{q} = \frac{1}{q!} \prod_{i=0}^{q-1}(K_n - i) \geq \frac{1}{q!} (K_n - q)^q$,  we use (\ref{eqn-Perfectmatchings-2})  to derive
\begin{align}
 & K_n \leq q + \sqrt[q]{q!} \cdot \sqrt[5q]{\frac{1}{q+2}\cdot \frac{(\ln n)^{1-\gamma}}{(\ln \ln n)^2}} \nonumber \\ & \text{for a constant }\gamma \in (0,1), \label{eqn-Perfectmatchings-3}
\end{align}
which along with the fact that $q$ is a constant further induces
\begin{align}
K_n  & = O\left(\sqrt[5q]{ \frac{(\ln n)^{1-\gamma}}{(\ln \ln n)^2}}\right) = o\left((\ln n)^{\frac{1}{5q}}\right). \label{eqn-Perfectmatchings-4}
\end{align}
To ensure   connectivity of a secure sensor network resulted from the $q$-composite key predistribution scheme, with $p_q$ denoting the edge probability, both \cite{Perfectmatchings} and our work obtain that it is necessary to have $p_q = \Omega\big(\frac{\ln n}{n}\big)$ or equivalently $\frac{1}{q!}
 \cdot \frac{{K_n}^{2q}}{{P_n} ^{q}}  = \Omega\big(\frac{\ln n}{n}\big)$, since it holds that $\frac{1}{q!}
 \cdot \frac{{K_n}^{2q}}{{P_n} ^{q}} $ is an   asymptotic expression of $p_q$ from (\ref{pssimq}) later. The requirement $\frac{1}{q!}
 \cdot \frac{{K_n}^{2q}}{{P_n} ^{q}}  = \Omega\big(\frac{\ln n}{n}\big)$ and the condition  (\ref{eqn-Perfectmatchings-4}) derived from  \cite{Perfectmatchings} together imply
 \begin{align}
P_n  & = \frac{{K_n}^2}{\sqrt[q]{q!\times \Omega\big(\frac{\ln n}{n}\big)}} = o\left(n^{\frac{1}{q}}(\ln n)^{-\frac{3}{5q}}\right). \label{eqn-Perfectmatchings-5}
\end{align}
However, we explain that (\ref{eqn-Perfectmatchings-5})
is not applicable to practical sensor networks. This paper for general $q$ and Di Pietro \emph{et al.}
\cite{4198829,DiPietroTissec} for $q=1$ have both proved that the condition $\frac{P_n}{K_n} =
\Omega(n)$ is needed to ensure reasonable network resiliency against node-capture attacks in the sense that an adversary
capturing sensors at random has to obtain at least a constant
fraction of nodes of the network in order to compromise a constant
fraction of secure links. The condition $\frac{P_n}{K_n} =
\Omega(n)$ with $K_n \geq 1$ clearly implies $P_n =
\Omega(n)$, which does not hold for any $P_n$ satisfying  (\ref{eqn-Perfectmatchings-5}). Hence, we have explained that the result of \cite{Perfectmatchings} cannot be used to design secure sensor networks in practice. In contrast, from condition set $\Lambda$ in (\ref{Lambda-set}), our theorems apply to a more practical range of $P_n = \omega(n \ln n)$. Hence, our results apply to real-world secure sensor networks and provide useful design guidelines.

For connectivity analysis of secure sensor networks taking into transmission constraints, for the EG scheme, Krzywdzi\'{n}ski and Rybarczyk
\cite{Krzywdzi}, and Krishnan \emph{et al.} \cite{ISIT_RKGRGG} recently show upper bounds $\frac{8\ln n}{n}$ and $\frac{2\pi \ln n}{n}$ for the critical threshold that the edge probability (i.e., the probability of a secure link between two sensors) takes for connectivity, while \cite{MobiCom14} proves   the {exact} threshold as $\frac{\ln n}{n}$. For the $q$-composite scheme with {general} $q$, \cite{JZNodeCapture} has presented connectivity results under transmission constraints. However, the following issues addressed in this paper are not tackled by  \cite{JZNodeCapture}:  the boundary effect of the network fields, the mobility of sensors, the combination of link unreliability and the transmission constraints modeled by the disk model (to be explained in Section \ref{secDiskModel}). We note that there are many connectivity studies \cite{yagan_onoff,yagan,yavuz2015toward,yavuz2017k,yagan2013modeling,IEEE-TSIPN-17} considering only link unreliability without the disk model (node mobility and node-capture attacks are not considered in these references).

\section{System Models} \label{sysmodel}

All of our studied networks employ the $q$-composite scheme.
Clearly, our analysis also applies to the basic Eschenauer--Gligor
scheme which has $q=1$. We first discuss the communication models
and mobility, and then detail the studied networks.

 \subsection{Communication Models}

We consider the following communication models.

 \subsubsection{Full Visibility}~

 The full visibility model
assumes that any two sensors have a direct communication link in
between. Under such model, a secure link exists between two sensors
if and only if they share at least $q$ keys.

 \subsubsection{Disk Model} \label{secDiskModel} ~

 In the disk model, each
node's transmission area is a disk with a transmission radius $r_n$,
with $r_n$ being a function of $n$ for generality. Two nodes have to
be within $r_n$ (their distance is at most $r_n$) for direct
communication. As for the node distribution, the same as much
previous work
\cite{ISIT_RKGRGG,Krzywdzi,Gupta98criticalpower,liu2016network},
we consider that the $n$ nodes
 are independently and uniformly deployed in some network region $\mathcal {A}$.
We let the region $\mathcal {A}$ be either a torus $\mathcal {T}$ or
a square $\mathcal {S}$, each with a unit area. The unit torus
$\mathcal{T}$
 eliminates the boundary effect (a node on $\mathcal{T}$ ``exits''
the area from one side appears as reentering from the opposite
side). The boundary effect of the unit square $\mathcal{S}$ is that a
circle with radius $r_n$ centered a point near the boundary of
$\mathcal{S}$ may have a part falling outside of $\mathcal{S}$, so a
node close to one side and another node close to the opposite side
may not have an edge in between on the square $\mathcal{S}$, but may
have an edge in between on the torus $\mathcal{T}$ because of wrap-around connections of
torus topology.

 \subsubsection{Link Unreliability}~

Communication links between nodes may not be available due to the
presence of physical barriers between nodes or because of harsh
environmental conditions severely impairing transmission. To model
unreliable links, each link is either {\em active} with probability
$t_n$ or {\em inactive} with probability $(1-t_n)$, where $t_n$ is a
function of $n$ for generality.

Table \ref{tabsym} summarizes the symbols and their meanings.

 \subsection{Mobility}

In addition to static sensor networks, we also consider mobile
sensor networks which are initialized in the same way as static
sensor networks. After initialization, sensors may move around. On
either the torus or the square, we consider the i.i.d. mobility model, where each node independently picks a new location uniformly at random at the beginning of a time slot and stays at the location in the rest of the time slot. This i.i.d. mobility model clearly   yields a uniform node distribution at each time
slot. For each mobile network that we consider, given a uniform node
distribution at each time slot, we can view the mobile network at a
single time slot as an instance of the corresponding static network. A future work is to consider other mobility models \cite{tomasini2013evaluating,xie2014survey,la2012network,liu2017information}.

 \subsection{Studied Networks}

We summarize our studied networks in Table \ref{table-networks}. As presented, we consider the following nine secure sensor networks:
\begin{itemize}
\item a secure sensor network with the $q$-composite scheme under full visibility (the network can be either static or mobile),
\item a static secure sensor network with the $q$-composite scheme under the disk model on the unit torus without the boundary effect,
\item a static secure sensor network with the $q$-composite scheme under the disk model on the unit square with the boundary effect,
\item a static secure sensor network with the $q$-composite scheme and unreliable links under the disk model on the unit torus without the boundary effect,
\item a static secure sensor network with the $q$-composite scheme and unreliable links under the disk model on the unit square with the boundary effect,
\item a mobile secure sensor network with the $q$-composite scheme under the disk model on the unit torus without the boundary effect,
\item a mobile secure sensor network with the $q$-composite scheme under the disk model on the unit square with the boundary effect,
\item a mobile secure sensor network with the $q$-composite scheme and unreliable links under the disk model on the unit torus without the boundary effect, and
\item a mobile secure sensor network with the $q$-composite scheme and unreliable links under the disk model on the unit square with the boundary effect.
\end{itemize}

\begin{table}[t]
\begin{center}
\begin{tabular}{|l|l|l|l|}
\hline \multicolumn{3}{|l|}{\textbf{Settings}} & \textbf{Notation
for networks}
\\ \hline \multicolumn{3}{|l|}{\begin{tabular}[c]{@{}l@{}}full
visibility\\ (static/mobile secure sensor networks)\end{tabular}}
& $\textnormal{network}^{\textnormal{full visibility}}$
\\ \hline \multirow{4}{*}{\begin{tabular}[c]{@{}l@{}}
\\ static\\secure\\sensor\\networks\end{tabular}}          & \multirow{2}{*}{disk model}                                                               & torus  & $\textnormal{static-network}^{\textnormal{disk model}}_{\textnormal{unit torus}}$                                 \\ \cline{3-4}
                                                                                             &                                                                                           & square & $\textnormal{static-network}^{\textnormal{disk model}}_{\textnormal{unit square}}$                                \\ \cline{2-4}
                                                                                             & \multirow{2}{*}{\begin{tabular}[c]{@{}l@{}}disk model w/\\ unreliable links\end{tabular}} & torus  & $\textnormal{static-network}^{\begin{subarray}\textnormal{\textnormal{disk model}}\\ \textnormal{unit torus}\end{subarray}}_{\textnormal{unreliable links}}$  \\ \cline{3-4}
                                                                                             &                                                                                           & square & $\textnormal{static-network}^{\begin{subarray}\textnormal{\textnormal{disk model}}\\ \textnormal{unit square}\end{subarray}}_{\textnormal{unreliable links}}$ \\ \hline
\multirow{4}{*}{\begin{tabular}[c]{@{}l@{}}
\\ mobile\\secure\\sensor\\networks\end{tabular}} & \multirow{2}{*}{disk model}                                                               & torus  & $\textnormal{mobile-network}^{\textnormal{disk model}}_{\textnormal{unit torus}}$                                 \\ \cline{3-4}
                                                                                             &                                                                                           & square & $\textnormal{mobile-network}^{\textnormal{disk model}}_{\textnormal{unit square}}$                                \\ \cline{2-4}
                                                                                             & \multirow{2}{*}{\begin{tabular}[c]{@{}l@{}}disk model w/\\ unreliable links\end{tabular}} & torus  & $\textnormal{mobile-network}^{\begin{subarray}\textnormal{\textnormal{disk model}}\\ \textnormal{unit torus}\end{subarray}}_{\textnormal{unreliable links}}$  \\ \cline{3-4}
                                                                                             &                                                                                           & square & $\textnormal{mobile-network}^{\begin{subarray}\textnormal{\textnormal{disk model}}\\ \textnormal{unit square}\end{subarray}}_{\textnormal{unreliable links}}$ \\ \hline
\end{tabular}
\end{center}
\caption{The studied secure sensor networks and their notation, where ``w/'' is short for ``with''. All networks employ the $q$-composite random key
predistribution scheme.}   \label{table-networks}
\end{table}

Throughout the paper, $q$ is an arbitrary positive integer and does
not scale with $n$, the number of nodes in the sensor network.  In
addition, $\ln$ is the natural logarithm function, the base of which
is $e$.

\section{Main Results} \label{sec:main:res}

We present the main results below.
\subsection{{Connectivity} of Secure Sensor Networks}
\label{sec-ding}

We establish connectivity results for the nine settings of secure
sensor networks in Table \ref{table-networks}. A network is
connected if each node can find at least one path to securely communicate with another
node. The results are summarized in Table \ref{tb-all-res}, and they
will be presented as theorems in detail.

\begin{table*}[t]
{\fontsize{8.5}{9}\selectfont\begin{tabular}{|l|l|l|l|}
\hline \multicolumn{3}{|l|}{ \hspace{-2pt}\textbf{Settings}
\hspace{-2pt}} & \hspace{-2pt}\textbf{Connectivity results}
\\ \hline \multicolumn{3}{|l|}{ \begin{tabular}[c]{@{}l@{}}\hspace{-2pt}static/mobile
secure sensor networks\hspace{-2pt}\\
 \hspace{-2pt}under full visibility\hspace{-2pt}\end{tabular}} &
\hspace{-2pt}connected if $\frac{1}{q!}  \cdot
\frac{{K_n}^{2q}}{{P_n} ^{q}} \geq \frac{c\ln n}{n}$ for any
constant $c>1$ \hspace{-2pt}
\\ \hline \multirow{4}{*}{\begin{tabular}[c]{@{}l@{}}
  \\ \hspace{-2pt}static\hspace{-2pt}\\
\hspace{-2pt}secure\hspace{-2pt}\\
\hspace{-2pt}sensor\hspace{-2pt}\\
\hspace{-2pt}networks\hspace{-20pt}\end{tabular}} &
\multirow{2}{*}{\hspace{-2pt}disk model \hspace{-2pt}} &
\hspace{-2pt}torus \hspace{-2pt}  &  \hspace{-2pt}connected if
$\frac{1}{q!}  \cdot \frac{{K_n}^{2q}}{{P_n} ^{q}}\cdot \pi{r_n}^{2}
\geq \frac{c\ln n}{n}$ for any constant $c>1$ \hspace{-2pt}
\\ \cline{3-4}
                                                                                    &                                                                                           &  \hspace{-2pt}square\hspace{-20pt} &  \hspace{-2pt}connected if $\frac{1}{q!}  \cdot \frac{{K_n}^{2q}}{{P_n} ^{q}} \cdot \pi{r_n}^{2} \geq \frac{c\ln n}{n}$ for any constant $c>c_n^{*}$ \hspace{-2pt}                                                                                                                         \\ \cline{2-4}
                                                                                    & \multirow{2}{*}{\begin{tabular}[c]{@{}l@{}} \hspace{-2pt}disk model w/\hspace{-2pt}\\  \hspace{-2pt}unreliable links\hspace{-20pt}\end{tabular}} &  \hspace{-2pt}torus\hspace{-20pt}  &  \hspace{-2pt}connected if $\frac{1}{q!}  \cdot \frac{{K_n}^{2q}}{{P_n} ^{q}}\cdot \pi{r_n}^{2} \cdot t_n \geq \frac{c\ln n}{n}$ for any constant $c>1$                                                                                                                      \\ \cline{3-4}
                                                                                    &                                                                                           &  \hspace{-2pt}square\hspace{-20pt} &  \hspace{-2pt}connected if $\frac{1}{q!}  \cdot \frac{{K_n}^{2q}}{{P_n} ^{q}} \cdot \pi{r_n}^{2} \cdot t_n \geq \frac{c\ln n}{n}$ for any constant $c>c_n^{\#}$                                                                                                              \\ \hline
\multirow{4}{*}{\begin{tabular}[c]{@{}l@{}}
\\ \hspace{-2pt}mobile\hspace{-2pt}\\
\hspace{-2pt}secure\hspace{-2pt}\\  \hspace{-2pt}sensor\\
\hspace{-2pt}networks\hspace{-20pt}\end{tabular}} &
\multirow{2}{*}{\hspace{-2pt}disk model\hspace{-2pt}} &
\hspace{-2pt}torus\hspace{-20pt}  &  \hspace{-2pt}connected for at
least $n^{c-1-\epsilon}$ consecutive time slots if $\frac{1}{q!}
 \cdot \frac{{K_n}^{2q}}{{P_n} ^{q}}\cdot \pi{r_n}^{2} \geq
\frac{c\ln n}{n}$ for any constant $c>1$                          \\
\cline{3-4}
                                                                                    &                                                                                           &  \hspace{-2pt}square\hspace{-20pt} &  \hspace{-2pt}connected for at least $n^{c-c_n^{*}-\epsilon}$ consecutive time slots if $\frac{1}{q!}  \cdot \frac{{K_n}^{2q}}{{P_n} ^{q}} \cdot \pi{r_n}^{2} \geq \frac{c\ln n}{n}$ for any constant $c>c_n^{*}$             \\ \cline{2-4}
                                                                                    & \multirow{2}{*}{\begin{tabular}[c]{@{}l@{}} \hspace{-2pt}disk model w/\hspace{-2pt}\\  \hspace{-2pt}unreliable links\hspace{-20pt}\end{tabular}} &  \hspace{-2pt}torus\hspace{-20pt}  &  \hspace{-2pt}connected for at least $n^{c-1-\epsilon}$ consecutive time slots if $\frac{1}{q!}  \cdot \frac{{K_n}^{2q}}{{P_n} ^{q}}\cdot \pi{r_n}^{2} \cdot t_n \geq \frac{c\ln n}{n}$ for any constant $c>1$                \\ \cline{3-4}
                                                                                    &                                                                                           &  \hspace{-2pt}square\hspace{-20pt} &  \hspace{-2pt}connected for at least $n^{c-c_n^{\#}-\epsilon}$ consecutive time slots if $\frac{1}{q!}  \cdot \frac{{K_n}^{2q}}{{P_n} ^{q}} \cdot \pi{r_n}^{2} \cdot t_n \geq \frac{c\ln n}{n}$ for any constant $c>c_n^{\#}$\hspace{-4pt} \\ \hline
\end{tabular}} \vspace{2pt}
  \caption{Different settings of secure sensor networks and their connectivity results, where ``w/'' is short for ``with''. Note that $c_n^{*} $ is $ \max \Big\{1 +  \big( \ln
\frac{P_n}{{K_n}^2} \big) \big/ \ln n , ~4 \big( \ln
\frac{P_n}{{K_n}^2} \big)\big/ \ln n  \Big\} $, $c_n^{\#}$ is $c_n^{\#}$ is $ \max \Big\{1 + \big( q \ln \frac{P_n}{{K_n}^2} + \ln \frac{1}{t_n} \big) \big/ \ln n , ~4 \big( q \ln \frac{P_n}{{K_n}^2} + \ln \frac{1}{t_n} \big) \big/ \ln n \Big \} $, {\color{white}and
$\epsilon$ is an arbitrary positive constant. All results are in the
asymptotic sense to hold with high probability, and thus the
conditions in the table only need to hold for all $n$ sufficiently
large. Except the result under full visibility which requires
conditions on $K_n$ and $P_n$ in condition set $\Lambda$ in (\ref{Lambda-set}) on Page \pageref{Lambda-set} but not the
condition $r_n=o(1)$ in $\Lambda$, all other results need the whole
condition set $\Lambda$.}}    \label{tb-all-res}
\end{table*}

%

All results given in Table \ref{tb-all-res} rely a set $\Lambda$ of
conditions, defined as follows:
%
%
%
\begin{align}
\textnormal{Condition set }\Lambda: \begin{cases} K_n& \hspace{-10pt} = \omega(\ln n),  \\
 K_n \hspace{1pt} &  \hspace{-10pt} = o\Big(\min\big\{\sqrt{P_n}, \frac{P_n}{n}\big\}\Big),   \\
\hspace{2pt} r_n &  \hspace{-10pt}= o(1) . \end{cases} \label{Lambda-set}
\end{align}

We explain that all conditions in set $\Lambda$ are practical. Note
that $K_n$ is the number of keys assigned to each sensor before
deployment. In real-world implementations, $K_n$ is often larger
\cite{yagan,ToN17-topological} than $\ln n$, so $K_n = \omega(\ln
n)$ follows. As concrete examples, we have $\ln 1000 \approx 6.9$,
$\ln 5000 \approx 8.5$ and $\ln 10000 \approx 9.2$. Since $K_n$ is
much smaller compared to both $n$ and $P_n$ due to constrained
memory and computational resources of sensors
\cite{virgil,adrian,yagan}, then $\frac{{K_n}^2}{P_n} = o(1)$
and $\frac{K_n}{P_n} = o\big(\frac{1}{n}\big)$ are practical,
yielding $K_n = o\big(\min\{\sqrt{P_n}, \frac{P_n}{n}\}\big)$. Note
that $K_n = \omega(\ln n)$ and $K_n = o\big(\min\{\sqrt{P_n},
\frac{P_n}{n}\}\big)$ together imply $P_n = \omega (n \ln n)$, which
is also practical since $P_n$ is larger
 than $n$ \cite{virgil,adrian,DiPietroTissec}. As examples, we have $1000 \ln 1000
\approx 6907$, $ 5000\ln 5000 \approx 42585$ and $10000\ln 10000
\approx 92103$. Finally, because we consider a unit area of network
region and there are $n$ nodes in this region, the condition $r_n =
o(1)$ is also practical.

To look at the conditions related to $\frac{\ln n}{n}$ in Table
\ref{tb-all-res}, we first explain that for each network in Table
\ref{tb-all-res}, the left hand side of each condition related to
$\frac{\ln n}{n}$ in the result
 is asymptotically equivalent to the
edge probability of the network, where the edge probability is the
probability that two nodes have an edge in between for direct
communication, and is defined for each time slot for mobile networks
(note that all nodes are symmetric in each network). Specifically,
we have the following under condition set $\Lambda$: \label{edge-prob-asymptotic-list}
\begin{itemize}
  \item[\ding {172}] $\frac{1}{q!}  \cdot
\frac{{K_n}^{2q}}{{P_n} ^{q}}$ is asymptotically equivalent to the
edge probability of $\textnormal{network}^{\textnormal{full
visibility}}$,
  \item[\ding {173}] $\frac{1}{q!}  \cdot \frac{{K_n}^{2q}}{{P_n} ^{q}}\cdot \pi{r_n}^{2}$ is a common
asymptotic value of the edge probabilities of networks
$\textnormal{static-network}^{\textnormal{disk
model}}_{\textnormal{unit torus}}$,
$\textnormal{static-network}^{\textnormal{disk
model}}_{\textnormal{unit square}}$,
$\textnormal{mobile-network}^{\textnormal{disk
model}}_{\textnormal{unit torus}}$, and
$\textnormal{mobile-network}^{\textnormal{disk
model}}_{\textnormal{unit square}}$, and
  \item[\ding {174}] $\frac{1}{q!}  \cdot \frac{{K_n}^{2q}}{{P_n} ^{q}}\cdot \pi{r_n}^{2} \cdot t_n$ is a common
asymptotic value of the edge probabilities of
$\textnormal{static-network}^{\begin{subarray}\textnormal{\textnormal{disk
model}}\\ \textnormal{unit
torus}\end{subarray}}_{\textnormal{unreliable links}}$,
$\textnormal{static-network}^{\begin{subarray}\textnormal{\textnormal{disk
model}}\\ \textnormal{unit
square}\end{subarray}}_{\textnormal{unreliable links}}$,
$\textnormal{mobile-network}^{\begin{subarray}\textnormal{\textnormal{disk
model}}\\ \textnormal{unit
torus}\end{subarray}}_{\textnormal{unreliable links}}$, and
$\textnormal{mobile-network}^{\begin{subarray}\textnormal{\textnormal{disk
model}}\\ \textnormal{unit
square}\end{subarray}}_{\textnormal{unreliable links}}$.
\end{itemize}

Before explaining the above results \ding {172}--\ding {174}, we first present a useful result (i.e., Lemma \ref{lem-rgg-edge-prob}) and its proof.

\begin{lem} \label{lem-rgg-edge-prob}
Let $\mu_n(\mathcal{A})$ be the probability that two nodes are within distance
$r_n$ when they are independently and uniformly distributed in a network region $\mathcal {A}$. Then we have:
\begin{itemize}
\item When $\mathcal {A}$ is a torus $\mathcal {T}$ of unit area, $\mu_n(\mathcal{A})$ becomes $\mu_n(\mathcal{T})$, which is given by
\begin{align}
\textstyle{\mu_n(\mathcal{T})  = \pi{r_n}^{2}, \text{ for } r_n \leq \frac{1}{2}} . \label{rgg-edge-prob-torus}
\end{align}
\item When $\mathcal {A}$ is a square $\mathcal {S}$ of unit area, $\mu_n(\mathcal{A})$ becomes $\mu_n(\mathcal{S})$, which satisfies
\begin{align}
\textstyle{ (1 - 2 {r_n})^2
\cdot \pi {r_n}^2  \leq \mu_n(\mathcal{S})  \leq \pi{r_n}^{2}, \text{ for } r_n \leq \frac{1}{2}},\label{rgg-edge-prob-square-1}
\end{align}
and
\begin{align}
\textstyle{\mu_n(\mathcal{S}) \sim \pi{r_n}^{2}, \text{ for } r_n = o(1)} , \label{rgg-edge-prob-square-2}
\end{align}
\end{itemize}
where $\mu_n(\mathcal{S}) \sim \pi{r_n}^{2}$ means $\lim_{n \to \infty} \frac{\mu_n(\mathcal{S}) }{\pi{r_n}^{2}} = 1$ (i.e., $\mu_n(\mathcal{S})$ is asymptotically equivalent to $\pi{r_n}^{2}$), and $r_n = o(1)$ means $\lim_{n \to \infty}r_n =0$.
\end{lem}

\noindent\textbf{Proof of Lemma \ref{lem-rgg-edge-prob}:}

Our goal is to analyze $\mu_n(\mathcal{A})$, which denotes  the probability that two nodes are within distance
$r_n$ when they are independently and uniformly deployed in a network region $\mathcal {A}$. We let $v_x$ and $v_y$  denote the two nodes, and write their distance as $\text{distance}(v_x, v_y)$.

\textbf{Proving (\ref{rgg-edge-prob-torus}) for $\mu_n(\mathcal{T}) $ on the unit \textit{torus} $\mathcal{T}$:} We consider the case of the unit \textit{torus} $\mathcal{T}$ here. Note that $\mu_n(\mathcal{T})$ means $\bP{\text{distance}(v_x, v_y)\leq r_n}$ when nodes $v_x$ and $v_y$ are independently and uniformly deployed in the unit torus $\mathcal{T}$. \textit{Given} any  location of node $v_x$, the event $\text{distance}(v_x, v_y)\leq r_n$ happens if and only if the random position of node $v_y$ is located in the circle centered at node $v_x$'s position with a radius of $r_n$; this clearly happens with probability $\pi{r_n}^{2}$ since the above circle is completely inside the torus $\mathcal{T}$ due to the following three reasons:
\begin{itemize}
\item we consider $ r_n \leq \frac{1}{2}$ in (\ref{rgg-edge-prob-torus});
\item the torus $\mathcal{T}$ has no boundary (if we think the torus as a square, then a point ``exits''
the torus area from one side appears as reentering from the opposite
side),
\item the circle centered at node $v_x$'s position with a radius of $r_n$ has an area of $\pi{r_n}^{2}$, and the torus $\mathcal{T}$ has an unit area.
\end{itemize}
Therefore, {given} any  location of $v_x$, the event $\text{distance}(v_x, v_y)\leq r_n$ on the unit torus $\mathcal{T}$ occurs with probability $\pi{r_n}^{2}$ when node $v_y$ is uniformly deployed in the unit torus $\mathcal{T}$. Now we consider that node $v_x$ is also uniformly distributed in the unit torus $\mathcal{T}$, and clearly $\text{distance}(v_x, v_y)\leq r_n$ still happens with probability $\pi{r_n}^{2}$. Hence, we have proved $\mu_n(\mathcal{T})  = \pi{r_n}^{2}, \text{ for } r_n \leq \frac{1}{2}$; i.e., (\ref{rgg-edge-prob-torus}) holds.

\textbf{Proving (\ref{rgg-edge-prob-square-1}) for $\mu_n(\mathcal{S}) $ on the unit square $\mathcal{S}$:} We consider the case of the unit \textit{square} $\mathcal{S}$ here. Note that $\mu_n(\mathcal{S})$ means $\bP{\text{distance}(v_x, v_y)\leq r_n}$ when nodes $v_x$ and $v_y$ are independently and uniformly deployed in the unit torus $\mathcal{S}$. {Given} any  location of $v_x$, the event $\text{distance}(v_x, v_y)\leq r_n$ happens if and only if the random position of node $v_y$ is located in the circle centered at node $v_x$'s position with a radius of $r_n$.

\begin{figure*}[!t]
\begin{center}
  \includegraphics[scale=0.6]{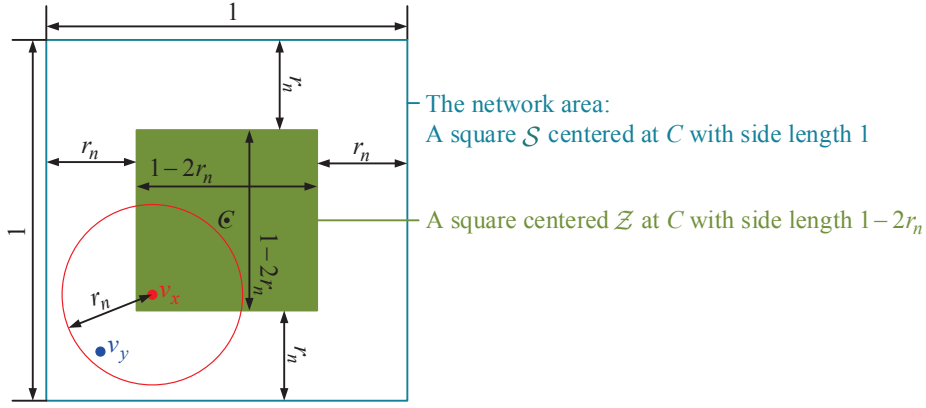}
  \caption{For the position of a node $v_x$
satisfying the condition that $v_x$'s distance to each edge of the
unit square is at least $ r_n$ with $r_n \leq
\frac{1}{2}$, given the position of $v_x$, the conditional probability that
another node $v_y$ falls in $v_x$'s transmission area is $\pi {r_n}^2$, since the circle centered at node $v_x$'s position with a radius of $r_n$ is completely inside the square.}   \label{fig-square}
  \end{center}
  \end{figure*}

We let $C$ denote the center of the unit square $\mathcal{S}$. Then for $ r_n \leq \frac{1}{2}$, we define $\mathcal{Z}$ as a square centered at $C$ with side length $1-2r_n$, as illustrated in Figure \ref{fig-square}. If node $v_x$'s location is inside $\mathcal{Z}$, we denote it by $v_x \in \mathcal{Z}$.

On the one hand, whenever $v_x \in \mathcal{Z}$, the circle centered at node $v_x$'s position with a radius of $r_n$ is completely inside the square $\mathcal{S}$, as shown in Figure \ref{fig-square}, so the random position of node $v_y$ belongs to this circle with probability $\pi{r_n}^{2}$. Hence, {given} any  location of $v_x$ which is inside $\mathcal{Z}$, the conditional probability that $v_x$ and $v_y$ is $\pi{r_n}^{2}$, when we consider the uniform distribution of
$v_y$'s position. Then the conditional probability of $\text{distance}(v_x, v_y) \leq r_n$ given $v_x \in \mathcal{Z}$ is $ \pi{r_n}^{2}$, for $r_n \leq \frac{1}{2}$. Formally, we obtain
\begin{align}
\textstyle{
 \bP{ \big(\text{distance}(v_x, v_y) \leq r_n \big) \boldsymbol{\mid}  (v_x \in \mathcal{Z} )}  = \pi{r_n}^{2}, \text{ for } r_n \leq \frac{1}{2}}.\label{rgg-edge-prob-square-3}
\end{align}

On the other hand, whenever $v_x \not \in \mathcal{Z}$, the circle centered at node $v_x$'s position with a radius of $r_n$ may or may not be completely inside the square $\mathcal{S}$, so the probability that the random position of node $v_y$ belongs to the intersection of this circle and the square $\mathcal{S}$ is no greater than $\pi{r_n}^{2}$. Hence, {given} any  location of $v_x$ which is outside of $\mathcal{Z}$, the conditional probability that $v_x$ and $v_y$ is at most $\pi{r_n}^{2}$, when we consider the uniform distribution of
$v_y$'s position.  Then the conditional probability of $\text{distance}(v_x, v_y) \leq r_n$ given $v_x \not \in \mathcal{Z}$ is no greater than $ \pi{r_n}^{2}$, for $r_n \leq \frac{1}{2}$. Formally, we have
\begin{align}
\textstyle{
 \bP{ \big(\text{distance}(v_x, v_y) \leq r_n \big) \boldsymbol{\mid}  (v_x \not \in \mathcal{Z} )} \leq \pi{r_n}^{2}, \text{ for } r_n \leq \frac{1}{2}}.\label{rgg-edge-prob-square-4}
\end{align}

Recall that $\mu_n(\mathcal{S}) $ denotes $\text{distance}(v_x, v_y) \leq r_n$ on the unit square $\mathcal{S}$. Clearly, from the law of total probability, we derive
\begin{align}
 &\mu_n(\mathcal{S}) \nonumber \\ & =  \bP{ \text{distance}(v_x, v_y) \leq r_n} \nonumber \\ & = \bP{ \big(\text{distance}(v_x, v_y) \leq r_n \big) \bcap (v_x \in \mathcal{Z} )} \nonumber \\ & \quad + \bP{ \big(\text{distance}(v_x, v_y) \leq r_n \big) \bcap (v_x \not\in \mathcal{Z} )} \nonumber \\ & =  \bP{ \big(\text{distance}(v_x, v_y) \leq r_n \big) \boldsymbol{\mid}  (v_x  \in \mathcal{Z} )} \times  \bP{v_x  \in \mathcal{Z} }   \nonumber \\ & \quad +  \bP{ \big(\text{distance}(v_x, v_y) \leq r_n \big) \boldsymbol{\mid}  (v_x \not \in \mathcal{Z} )} \times  \bP{v_x \not \in \mathcal{Z} }  .\label{rgg-edge-prob-square-5}
\end{align}
Ignoring the term after the plus sign in (\ref{rgg-edge-prob-square-5}), we further have
\begin{align}
\mu_n(\mathcal{S}) & \geq \bP{ \big(\text{distance}(v_x, v_y) \leq r_n \big) \boldsymbol{\mid}  (v_x  \in \mathcal{Z} )} \times  \bP{v_x  \in \mathcal{Z} }   .\label{rgg-edge-prob-square-6}
\end{align}

Note that $v_x$ is uniformly distributed on the  square $\mathcal{S}$, whose area denoted by $|\mathcal{S}|$ is given by $|\mathcal{S}|=1$. The square $\mathcal{Z}$ has a side length of $1-2r_n$, so its area denoted by $|\mathcal{Z}|$ equals $|\mathcal{Z}|=(1 - 2 {r_n})^2$. Hence, we obtain
\begin{align}
\bP{v_x \in \mathcal{Z} } = \frac{|\mathcal{Z}|}{|\mathcal{S}|} = \frac{ (1 - 2 {r_n})^2}{1} =  (1 - 2 {r_n})^2, \text{ for } r_n \leq \textstyle{\frac{1}{2}}  ,\label{rgg-edge-prob-square-7}
\end{align}
and
\begin{align}
\bP{v_x \not\in \mathcal{Z} } = 1 - \bP{v_x \in \mathcal{Z} } = 1- (1 - 2 {r_n})^2, \text{ for } r_n \leq \textstyle{\frac{1}{2}}  .\label{rgg-edge-prob-square-8}
\end{align}

Using (\ref{rgg-edge-prob-square-3}) and (\ref{rgg-edge-prob-square-7}) in (\ref{rgg-edge-prob-square-6}), we get
\begin{align}
\mu_n(\mathcal{S}) & \geq  \pi{r_n}^{2} \times  (1 - 2 {r_n})^2, \text{ for } r_n \leq \textstyle{\frac{1}{2}}  . \label{rgg-edge-prob-square-9}
\end{align}
Applying (\ref{rgg-edge-prob-square-3}) (\ref{rgg-edge-prob-square-4}) (\ref{rgg-edge-prob-square-7}) and (\ref{rgg-edge-prob-square-8}) to (\ref{rgg-edge-prob-square-5}), we establish
\begin{align}
\mu_n(\mathcal{S}) & \leq  \pi{r_n}^{2} \times  (1 - 2 {r_n})^2 +  \pi{r_n}^{2} \times  [1-(1 - 2 {r_n})^2] \nonumber \\ & \leq  \pi{r_n}^{2}, \text{ for } r_n \leq \textstyle{\frac{1}{2}}  . \label{rgg-edge-prob-square-10}
\end{align}
Summarizing (\ref{rgg-edge-prob-square-9}) and (\ref{rgg-edge-prob-square-10}), we have
$ (1 - 2 {r_n})^2
\cdot \pi {r_n}^2  \leq \mu_n(\mathcal{S})  \leq \pi{r_n}^{2}, \text{ for } r_n \leq \frac{1}{2}$; i.e., (\ref{rgg-edge-prob-square-1}) is proved.



\textbf{Proving (\ref{rgg-edge-prob-square-2}) for $\mu_n(\mathcal{S}) $ on the unit square $\mathcal{S}$:} Recall that for two
positive sequences $x_n$ and $y_n$, the relation $x_n \sim y_n$
means $\lim_{n \to
  \infty} \frac{x_n}{y_n} =1$.   From (\ref{rgg-edge-prob-square-1}), it follows that $ (1 - 2 {r_n})^2 \leq \frac{\mu_n(\mathcal{S}) }{\pi{r_n}^{2}}  \leq 1, \text{ for } r_n \leq \frac{1}{2}$. Clearly, given $r_n = o(1)$, we have $\lim_{n \to
  \infty} (1 - 2 {r_n})^2 =1$ and also have $r_n \leq \frac{1}{2}$ for all $n$ sufficiently large. Summarizing the above results, we clearly obtain  $\lim_{n \to \infty} \frac{\mu_n(\mathcal{S}) }{\pi{r_n}^{2}} = 1$; i.e., $\mu_n(\mathcal{S}) \sim \pi{r_n}^{2}$. Hence, (\ref{rgg-edge-prob-square-2}) is proved.
\qeda







We now show results \ding {172}--\ding {174} of Page \pageref{edge-prob-asymptotic-list}.

We first show the above result \ding {172}. By \cite[Lemma
1]{JZISIT14}, if $K_n = \omega(1)$ and $K_n =
o\big(\sqrt{P_n}\big)$, then
\begin{align}
\textstyle{p_q  \sim \frac{1}{q!}
 \cdot \frac{{K_n}^{2q}}{{P_n} ^{q}}} .
\label{pssimq}
 \end{align}
The asymptotic equivalence (\ref{pssimq}) above holds for
$\textnormal{network}^{\textnormal{full visibility}}$ under
condition set $\Lambda$ in (\ref{Lambda-set}) on Page \pageref{Lambda-set}, since we have $K_n = \omega(1)$ and $K_n =
o\big(\sqrt{P_n}\big)$ from $\Lambda$.

We then explain the result \ding {173}. With
$\mu_n(\mathcal{T})$ (resp., $\mu_n(\mathcal{S})$) denoting the probability that two nodes are within distance
$r_n$ when they are independently and uniformly distributed on the unit torus $\mathcal {T}$ (resp., the unit square $\mathcal{S}$), Lemma \ref{lem-rgg-edge-prob} above shows $\textstyle{\mu_n(\mathcal{T})  = \pi{r_n}^{2}, \text{ for } r_n \leq \frac{1}{2}}$, and $\mu_n(\mathcal{S}) \sim \pi{r_n}^{2}, \text{ for } r_n = o(1)$. We can use Lemma \ref{lem-rgg-edge-prob}   because our condition set $\Lambda$ in (\ref{Lambda-set}) on Page \pageref{Lambda-set} has the condition $r_n = o(1)$, which implies $r_n \leq \frac{1}{2}$ for all $n$ sufficiently large. In words, Lemma \ref{lem-rgg-edge-prob} says that the probability that
two nodes are within distance $r_n$ is given by $\pi{r_n}^{2}$ on the unit torus $\mathcal {T}$, and is asymptotically equivalent to $\pi {r_n}^2$ on the unit square $\mathcal{S}$.
 In each of $\textnormal{static-network}^{\textnormal{disk
model}}_{\textnormal{unit torus}}$ and
$\textnormal{static-network}^{\textnormal{disk
model}}_{\textnormal{unit square}}$, since two nodes have an edge in
between if and only if they are within distance $r_n$ and also they
share at least $q$ keys, then under condition set $\Lambda$, the
edge probability is asymptotically equivalent to $\frac{1}{q!} \cdot
\frac{{K_n}^{2q}}{{P_n} ^{q}}\cdot \pi{r_n}^{2}$ from $ \mu_n(\mathcal{T})  = \pi{r_n}^{2}$, $ \mu_n(\mathcal{S})  \sim
\pi {r_n}^2  $ and $p_q  \sim \frac{1}{q!}
 \cdot \frac{{K_n}^{2q}}{{P_n} ^{q}}$ in (\ref{pssimq}).
 Furthermore, at each time slot, mobile networks $\textnormal{mobile-network}^{\textnormal{disk
model}}_{\textnormal{unit torus}}$ and
$\textnormal{mobile-network}^{\textnormal{disk
model}}_{\textnormal{unit square}}$ can be viewed as instances of
static networks $\textnormal{static-network}^{\textnormal{disk
model}}_{\textnormal{unit torus}}$ and
$\textnormal{static-network}^{\textnormal{disk
model}}_{\textnormal{unit square}}$, respectively. In view of the
above, the result \ding {173} is proved.

We now show the result \ding {174}. Compared with the networks in
the result \ding {173}, the networks in the result \ding {174} also
consider link unreliability, in which each link is allowed to be
inactive with probability $t_n$. Multiplying the term $\frac{1}{q!}
\cdot \frac{{K_n}^{2q}}{{P_n} ^{q}}\cdot \pi{r_n}^{2}$ with $t_n$,
we obtain $\frac{1}{q!} \cdot \frac{{K_n}^{2q}}{{P_n} ^{q}}\cdot
\pi{r_n}^{2} \cdot t_n $, which is a common asymptotic value of the
edge probabilities of the networks in the result \ding {174}.
Therefore, the result \ding {174} follows.

%


%
%
%
%
%
%

We have shown the results \ding {172} \ding {173} and \ding {174} on
the asymptotic values of the edge probabilities of the networks in
Table \ref{tb-all-res}. For each network in Table \ref{tb-all-res},
let $\widetilde{\textnormal{edge-prob}}$ denote the used asymptotic
value of its edge probability; namely,
$\widetilde{\textnormal{edge-prob}}$ is $\frac{1}{q!} \cdot
\frac{{K_n}^{2q}}{{P_n} ^{q}}$  for
$\textnormal{network}^{\textnormal{full visibility}}$ in the result
\ding {172}, is $\frac{1}{q!}  \cdot \frac{{K_n}^{2q}}{{P_n}
^{q}}\cdot \pi{r_n}^{2}$ for the four networks in the result \ding
{173}, and is $\frac{1}{q!}  \cdot \frac{{K_n}^{2q}}{{P_n}
^{q}}\cdot \pi{r_n}^{2} \cdot t_n$ for the four networks in the
result \ding {174}. As given in Table \ref{tb-all-res}, the
conditions related to $\frac{\ln n}{n}$ can be summarized as
follows. For $\textnormal{network}^{\textnormal{full visibility}}$
and the four networks on the unit torus, we have the condition that
$\widetilde{\textnormal{edge-prob}} \geq \frac{c\ln n}{n}$ for any
constant $c>1$. For the four networks on the unit square, we have
the condition that $\widetilde{\textnormal{edge-prob}} \geq
\frac{c\ln n}{n}$ for any constant $c> \widetilde{c}_n$, where
$\widetilde{c}_n$ is given by
\begin{align}
& \widetilde{c}_n :   = \max \bigg\{ 1 \hspace{-1pt}+ \hspace{-1pt}
\Big( \hspace{-2pt}\ln
\big[(\widetilde{\textnormal{edge-prob}})^{-1}\big]
\hspace{-2pt}\Big) \hspace{-2pt}\bigg/ \hspace{-2pt} \ln n,
\nonumber
\\  & \quad\quad\quad\quad\quad\quad\quad  4   \Big( \hspace{-2pt}\ln
\big[(\widetilde{\textnormal{edge-prob}})^{-1}\big]
\hspace{-2pt}\Big) \hspace{-2pt}\bigg/ \hspace{-2pt}\ln n
\hspace{-2pt} \bigg\} \nonumber \\  & \hspace{-2pt} = \hspace{-2pt}
\begin{cases} 1 \hspace{-2pt}+ \hspace{-2pt}
\Big( \hspace{-2pt}\ln
\big[(\widetilde{\textnormal{edge-prob}})^{-1}\big]
\hspace{-2pt}\Big) \hspace{-3pt}\bigg/ \hspace{-3pt} \ln n ,
\textnormal{ if } \widetilde{\textnormal{edge-prob}} \hspace{-2pt} =
\hspace{-2pt} \widetilde{\Omega}(n^{-1/3}), \\ 4   \Big(
\hspace{-2pt}\ln \big[(\widetilde{\textnormal{edge-prob}})^{-1}\big]
\Big) \hspace{-3pt}\bigg/ \hspace{-3pt}\ln n , \textnormal{ if }
\widetilde{\textnormal{edge-prob}}  \hspace{-2pt}= \hspace{-2pt}
\widetilde{O}(n^{-1/3}).
\end{cases} \label{zhao}
\end{align}

In (\ref{zhao}) above, we write a quantity $x_n$ as
$\widetilde{\Omega}(n^{\alpha})$ for some constant $\alpha$ if $x_n
\geq n^{\alpha-\epsilon_1}$ for all $n$ sufficiently large, where
$\epsilon_1$ is an arbitrary positive constant. Similarly, a
quantity $x_n$ is written as $\widetilde{O}(n^{\alpha})$ for some
constant $\alpha$ if $x_n \leq n^{\alpha+\epsilon_2}$ for all $n$
sufficiently large, where $\epsilon_2$ is an arbitrary positive
constant. 

The difference between the condition $c>1$ for networks on the torus, and
the condition $c>\widetilde{c}_n$  for networks on the square is resulted
from the boundary effect of the square.

From (\ref{zhao}), we observe a phase transition of
$\widetilde{c}_n$ when $\widetilde{\textnormal{edge-prob}}$ or the
edge probability asymptotically changes from
$\widetilde{\Omega}(n^{-1/3})$ to $\widetilde{O}(n^{-1/3})$, or vice
versa. The intuition is that to compute the expected number of
isolated nodes, in integrating all possible points in unit square
$\mathcal{S}$ with the boundary effect, different areas on
$\mathcal{S}$ contribute to the dominant part of the integral
depending on the edge probability, where a node is non-isolated if it has a link with at least another node, and a node is isolated if it has no link with any other node.

We now present the connectivity results in Table \ref{tb-all-res} as
more understandable theorems, which are all proved in Section
\ref{prove:thm}. We give
first the results of secure sensor networks under full visibility,
then the results of static secure sensor networks, and finally those
of mobile secure sensor networks. As mentioned before in Section
\ref{sysmodel} for the system model, all studied networks employ the
$q$-composite scheme.

\subsubsection{{Connectivity} of Secure Sensor Networks with the $q$-Composite
Scheme under Full Visibility}~
\label{sec:Connectivity:FullVisibility}


Theorem \ref{Connect-full-visi} below on connectivity is for the
full visibility case. Its proof is provided in Section
\ref{secprf-Connect-full-visi}.


\begin{thm}[Connectivity of $\textnormal{network}^{\textnormal{full visibility}}$] \label{Connect-full-visi}
Under condition set $\Lambda$ in (\ref{Lambda-set}) on Page \pageref{Lambda-set}, a secure sensor
network with the $q$-composite scheme under full visibility is
\emph{connected} with high probability if $\frac{1}{q!}
 \cdot \frac{{K_n}^{2q}}{{P_n} ^{q}}  \geq \frac{c\ln n}{n}$ for all $n$ sufficiently large,
 where $c$ is a constant with $c>1$. 

\end{thm}

As explained in the result \ding {172} in Section \ref{sec-ding},
$\frac{1}{q!}  \cdot \frac{{K_n}^{2q}}{{P_n} ^{q}}$ is an asymptotic
value of the edge probability of
$\textnormal{network}^{\textnormal{full visibility}}$. Although  Theorem \ref{Connect-full-visi} can also be derived using the results in \cite{ANALCO}, the proof techniques in  \cite{ANALCO} lack formality (specifically \cite[Equation (6.93)]{ANALCO}) and are different from those in this paper. In addition, Bloznelis and Rybarczyk
\cite{Bloznelis201494} (resp., Bloznelis and {\L}uczak \cite{Perfectmatchings}) have recently investigated
  connectivity (resp., $k$-connectivity) of $\textnormal{network}^{\textnormal{full visibility}}$, but both results after a rewriting \vspace{2pt} address only the narrow and impractical range of $P_n =o\big(n^{\frac{1}{q}} (\ln n)^{-\frac{3}{5q}}\big)$. In contrast, from condition set $\Lambda$ in (\ref{Lambda-set}), our Theorem \ref{Connect-full-visi} considers a more practical range of $P_n = \omega(n \ln n)$. More details can be found in Section
\ref{sec:related-work}  for related work.

\subsubsection{{Connectivity} of Static Secure Sensor Networks with the $q$-Composite
Scheme}~

Theorems \ref{static-torus}--\ref{static-square-unreliable} present
connectivity results of static secure sensor networks. Their proofs
are deferred to Section \ref{sec:static-torussquarepf}.

\paragraph{Connectivity under the Disk Model on the Torus without the
Boundary Effect}~

\begin{thm}[Connectivity of $\textnormal{static-network}^{\textnormal{disk model}}_{\textnormal{unit torus}}$] \label{static-torus}
Under condition set $\Lambda$ in (\ref{Lambda-set}) on Page \pageref{Lambda-set}, a static secure
sensor network with the $q$-composite scheme under the disk model on
the unit torus without the boundary effect is \emph{connected} with
high probability if $\frac{1}{q!}
 \cdot \frac{{K_n}^{2q}}{{P_n} ^{q}}  \cdot \pi {r_n}^{2} \geq \frac{c\ln n}{n}$ for all $n$ sufficiently large,
 where $c$ is a constant with $c>1$. 

\end{thm}

As given in the result \ding {173} in Section \ref{sec-ding},
\mbox{$\frac{1}{q!}
 \cdot \frac{{K_n}^{2q}}{{P_n} ^{q}}  \cdot \pi {r_n}^{2}$} is asymptotically equivalent to the edge probability
 of
$\textnormal{static-network}^{\textnormal{disk
model}}_{\textnormal{unit torus}}$.

\paragraph{Connectivity with Unreliable Links under the Disk Model on the Torus without the Boundary Effect}

\begin{thm}[Connectivity of $\textnormal{static-network}^{\begin{subarray}\textnormal{\textnormal{disk model}}\\ \textnormal{unit torus}\end{subarray}}_{\textnormal{unreliable links}}$] \label{static-torus-unreliable}

Under condition set $\Lambda$ in (\ref{Lambda-set}) on Page \pageref{Lambda-set}, a static secure
sensor network with the $q$-composite scheme and unreliable links
under the disk model on the unit torus without the boundary effect
is \emph{connected} with high probability if $\frac{1}{q!}
 \cdot \frac{{K_n}^{2q}}{{P_n} ^{q}}  \cdot \pi {r_n}^{2} \cdot t_n \geq
\frac{c\ln n}{n}$ for all $n$ sufficiently large,
 where $c$ is a constant with $c>1$.

\end{thm}

As explained in the result \ding {174} in Section \ref{sec-ding}, \mbox{$\frac{1}{q!}
 \cdot \frac{{K_n}^{2q}}{{P_n} ^{q}}  \cdot \pi {r_n}^{2} \cdot t_n $}
 is an
asymptotic value of the edge probability of
$\textnormal{static-network}^{\begin{subarray}\textnormal{\textnormal{disk
model}}\\ \textnormal{unit
torus}\end{subarray}}_{\textnormal{unreliable links}}$.

\paragraph{Connectivity under the Disk Model on the Square with the Boundary Effect}

\begin{thm}[Connectivity of $\textnormal{static-network}^{\textnormal{disk model}}_{\textnormal{unit square}}$] \label{static-square}

Under condition set $\Lambda$ in (\ref{Lambda-set}) on Page \pageref{Lambda-set}, a static secure
sensor network with the $q$-composite scheme under the disk model on
the unit square with the boundary effect is \emph{connected} with
high probability if $\frac{1}{q!}  \cdot \frac{{K_n}^{2q}}{{P_n}
^{q}}  \cdot \pi {r_n}^{2} \geq \frac{c\ln n}{n}$ for all $n$
sufficiently large, where $c$ is some constant satisfying for all
$n$ sufficiently large that
\begin{align}
c>  c_n^{*} :=\max \bigg\{ \hspace{-1pt}1 \hspace{-1pt}+
\hspace{-1pt} \bigg( \hspace{-2pt}\ln \frac{P_n}{{K_n}^2 }
\hspace{-2pt}\bigg) \hspace{-2pt}\bigg/ \hspace{-2pt} \ln n ,
\hspace{2pt}4  \bigg( \hspace{-2pt} \ln \frac{P_n}{{K_n}^2 }
\hspace{-2pt}\bigg) \hspace{-2pt}\bigg/ \hspace{-2pt}\ln n
\hspace{-2pt} \bigg\} . \label{cnstar}
\end{align}
\end{thm}

As given in the result \ding {173} in Section \ref{sec-ding}, \mbox{$\frac{1}{q!}
 \cdot \frac{{K_n}^{2q}}{{P_n} ^{q}}  \cdot \pi {r_n}^{2}$}
 is asymptotically equivalent to the edge probability
 of
$\textnormal{static-network}^{\textnormal{disk
model}}_{\textnormal{unit square}}$.

\paragraph{Connectivity with Unreliable Links under the Disk Model on the Square with the Boundary Effect}

\begin{thm}[Connectivity of $\textnormal{static-network}^{\begin{subarray}\textnormal{\textnormal{disk model}}\\ \textnormal{unit square}\end{subarray}}_{\textnormal{unreliable links}}$] \label{static-square-unreliable}

Under condition set $\Lambda$ in (\ref{Lambda-set}) on Page \pageref{Lambda-set}, a static secure
sensor network with the $q$-composite scheme and unreliable links
under the disk model on the unit square with the boundary effect is
\emph{connected} with high probability if $\frac{1}{q!}  \cdot
\frac{{K_n}^{2q}}{{P_n} ^{q}}  \cdot \pi {r_n}^{2} \cdot t_n \geq
\frac{c\ln n}{n}$ for all $n$ sufficiently large, where $c$ is some
constant satisfying for all $n$ sufficiently large that
\begin{align}
&c > c_n^{\#} := \max \bigg\{ 1 + \bigg( q \ln \frac{P_n}{{K_n}^2} + \ln \frac{1}{t_n} \bigg) \bigg/ \ln n , \nonumber
\\ & ~~~~~~~~~~~~~~~~~~~~~~~~~4 \bigg( q \ln \frac{P_n}{{K_n}^2} + \ln \frac{1}{t_n} \bigg) \bigg/ \ln n \bigg\} . \label{cnpound}
\end{align}

\end{thm}

As explained in the result \ding {174} in Section \ref{sec-ding}, \mbox{$\frac{1}{q!}
 \cdot \frac{{K_n}^{2q}}{{P_n} ^{q}}  \cdot \pi {r_n}^{2} \cdot t_n $}
 is an
asymptotic value of the edge probability of
$\textnormal{static-network}^{\begin{subarray}\textnormal{\textnormal{disk
model}}\\ \textnormal{unit
square}\end{subarray}}_{\textnormal{unreliable links}}$.

\subsubsection{Connectivity of Mobile Secure Sensor Networks with the $q$-Composite
Scheme}~

Theorems \ref{mobile-torus}--\ref{mobile-square-unreliable} below
present connectivity results of mobile secure sensor networks. Their
proofs are deferred to Section \ref{sec:mobile-torussquarepf}. As
explained in Section \ref{sysmodel} for the system model, at each
time slot, each mobile network can be viewed as an instance of its
corresponding static network, and thus its edge probability defined
for each time slot also equals that of the corresponding static
network.

\paragraph{Connectivity under the Disk Model on the Torus without the Boundary Effect}

\begin{thm}[Connectivity of $\textnormal{mobile-network}^{\textnormal{disk model}}_{\textnormal{unit
torus}}$]  \label{mobile-torus}

Under condition set $\Lambda$ in (\ref{Lambda-set}) on Page \pageref{Lambda-set}, a mobile secure sensor network with the $q$-composite scheme under the disk model on the unit torus without the boundary effect and under the i.i.d. mobility model is connected with high probability for at least $
n^{c-1-\epsilon}$ consecutive time slots from the beginning for an arbitrary positive constant $\epsilon$, if $\frac{1}{q!}  \cdot \frac{{K_n}^{2q}}{{P_n} ^{q}}  \cdot \pi
{r_n}^{2} \geq \frac{c\ln n}{n}$ for all $n$ sufficiently large,
 where $c$ is a constant with $c>1$.

\end{thm}

As given in the result \ding {173} in Section \ref{sec-ding}, \mbox{$\frac{1}{q!}
 \cdot \frac{{K_n}^{2q}}{{P_n} ^{q}}  \cdot \pi {r_n}^{2}$}
 is asymptotically equivalent to the edge probability
 of
$\textnormal{mobile-network}^{\textnormal{disk
model}}_{\textnormal{unit torus}}$. Also, from Theorem
\ref{static-torus}, the conditions in Theorem \ref{mobile-torus}
also ensure that the corresponding static network
$\textnormal{static-network}^{\textnormal{disk
model}}_{\textnormal{unit torus}}$ is connected with high
probability.

\paragraph{Connectivity with Unreliable Links under the Disk Model on the Torus without the Boundary Effect}

\begin{thm}[Connectivity of $\textnormal{mobile-network}^{\begin{subarray}\textnormal{\textnormal{disk model}}\\ \textnormal{unit torus}\end{subarray}}_{\textnormal{unreliable
links}}$] \label{mobile-torus-unreliable}

Under condition set $\Lambda$ in (\ref{Lambda-set}) on Page \pageref{Lambda-set}, a mobile secure sensor network with the $q$-composite scheme and unreliable links under the disk model on the unit torus without the boundary effect and under the i.i.d. mobility model is connected with high probability for at least $
n^{c-1-\epsilon}$ consecutive time slots from the beginning for an arbitrary positive constant $\epsilon$, if $\frac{1}{q!}  \cdot \frac{{K_n}^{2q}}{{P_n} ^{q}}  \cdot \pi
{r_n}^{2}  \cdot t_n \geq \frac{c\ln n}{n}$ for all $n$ sufficiently large,
 where $c$ is a constant with $c>1$.

\end{thm}

As explained in the result \ding {174} in Section \ref{sec-ding}, \mbox{$\frac{1}{q!}
 \cdot \frac{{K_n}^{2q}}{{P_n} ^{q}}  \cdot \pi {r_n}^{2} \cdot t_n $}
 is an
asymptotic value of the edge probability of
$\textnormal{mobile-network}^{\begin{subarray}\textnormal{\textnormal{disk
model}}\\ \textnormal{unit
torus}\end{subarray}}_{\textnormal{unreliable links}}$. Moreover,
from Theorem \ref{static-torus-unreliable}, the conditions in
Theorem \ref{mobile-torus-unreliable} also ensure that the
corresponding static network
$\textnormal{static-network}^{\begin{subarray}\textnormal{\textnormal{disk
model}}\\ \textnormal{unit
torus}\end{subarray}}_{\textnormal{unreliable links}}$ is connected
with high probability.

\paragraph{Connectivity under the Disk Model on the Square with the Boundary Effect}

\begin{thm}[Connectivity of $\textnormal{mobile-network}^{\textnormal{disk model}}_{\textnormal{unit square}}$] \label{mobile-square}

Under condition set $\Lambda$ in (\ref{Lambda-set}) on Page \pageref{Lambda-set}, a mobile secure
sensor network with the $q$-composite scheme under the disk model on
the unit square with the boundary effect and under {the i.i.d.
mobility model} is connected with high probability for at least $
n^{c-c_n^{*}-\epsilon}$ consecutive time slots from the beginning
for an arbitrary positive constant $\epsilon$, if $\frac{1}{q!}
\cdot \frac{{K_n}^{2q}}{{P_n} ^{q}}  \cdot \pi {r_n}^{2} \geq
\frac{c\ln n}{n}$ for all $n$ sufficiently large, where $c$ is some
constant satisfying $c> c_n^{*}$ for all $n$ sufficiently large,
with $c_n^{*}$ defined in (\ref{cnstar}).
\end{thm}

As given in the result \ding {173} in Section \ref{sec-ding}, \mbox{$\frac{1}{q!}
 \cdot \frac{{K_n}^{2q}}{{P_n} ^{q}}  \cdot \pi {r_n}^{2}$}
 is asymptotically equivalent to the edge probability
 of
$\textnormal{mobile-network}^{\textnormal{disk
model}}_{\textnormal{unit square}}$. In addition, from Theorem
\ref{static-square}, the conditions in Theorem \ref{mobile-square}
also ensure that the corresponding static network
$\textnormal{static-network}^{\textnormal{disk
model}}_{\textnormal{unit square}}$ is connected with high
probability.

\paragraph{Connectivity with Unreliable Links under the Disk Model on the Square with the Boundary Effect}

\begin{thm}[Connectivity of $\textnormal{mobile-network}^{\begin{subarray}\textnormal{\textnormal{disk model}}\\ \textnormal{unit square}\end{subarray}}_{\textnormal{unreliable
links}}$] \label{mobile-square-unreliable}

Under condition set $\Lambda$ in (\ref{Lambda-set}) on Page \pageref{Lambda-set}, a mobile secure
sensor network with the $q$-composite scheme and unreliable links
under the disk model on the unit square with the boundary effect and
under {the i.i.d. mobility model} is connected with high
probability for at least $ n^{c-c_n^{\#}-\epsilon}$ consecutive time
slots from the beginning for an arbitrary positive constant
$\epsilon$, if $\frac{1}{q!}  \cdot \frac{{K_n}^{2q}}{{P_n} ^{q}}
\cdot \pi {r_n}^{2}  \cdot t_n  \geq  \frac{c\ln n}{n}$ for all $n$
sufficiently large, where $c$ is some constant satisfying $c>
c_n^{\#}$ for all $n$ sufficiently large, with $c_n^{\#}$ defined in
(\ref{cnpound}).

\end{thm}

As explained in the result \ding {174} in Section \ref{sec-ding},
\mbox{$\frac{1}{q!}
 \cdot \frac{{K_n}^{2q}}{{P_n} ^{q}}  \cdot \pi {r_n}^{2} \cdot t_n $} is an
asymptotic value of the edge probability of
$\textnormal{mobile-network}^{\begin{subarray}\textnormal{\textnormal{disk
model}}\\ \textnormal{unit
square}\end{subarray}}_{\textnormal{unreliable links}}$.
Furthermore, from Theorem \ref{static-square-unreliable}, the
conditions in Theorem \ref{mobile-square-unreliable} also ensure
that the corresponding static network
$\textnormal{static-network}^{\begin{subarray}\textnormal{\textnormal{disk
model}}\\ \textnormal{unit
square}\end{subarray}}_{\textnormal{unreliable links}}$ is connected
with high probability.

\subsubsection{Connectivity under Random Node-Capture Attacks}

We present below the connectivity results under random node-capture
attacks. After the adversary has captured some \emph{random} set of
$m$ nodes, the remaining network is statistically equivalent to a
network initially with $(n-m)$ nodes in the absence of node capture.
Hence, by replacing $\ln n$ with $\ln (n-m)$, replacing $\frac{\ln
n}{n}$ with $\frac{\ln (n-m)}{n-m}$ and $n$ with $(n-m)$ in the
connectivity results in Table \ref{tb-all-res}, we obtain the
connectivity results in Table \ref{tb-all-res-new} for an $n$-size
network after the adversary has captured a random set of $m$ nodes
out of all $n$ nodes. Note that we do not replace the ``$n$'' in $K_n$, $P_n$, $r_n$, $t_n$.

\begin{table*}[t]
\hspace{-10pt}{\fontsize{8}{9}\selectfont\begin{tabular}{|l|l|l|l|}
\hline \multicolumn{3}{|l|}{ \hspace{-2pt}\textbf{Settings}
\hspace{-2pt}} & \hspace{-2pt}\textbf{Connectivity results under a
random capture of $m$ nodes}
\\ \hline \multicolumn{3}{|l|}{ \begin{tabular}[c]{@{}l@{}}\hspace{-2pt}static/mobile
secure sensor\hspace{-20pt}\\
 \hspace{-2pt}networks under full visibility\hspace{-30pt}\end{tabular}} &
\hspace{-2pt}connected if $\frac{1}{q!}  \cdot
\frac{{K_n}^{2q}}{{P_n} ^{q}} \geq \frac{c\ln (n-m)}{n-m}$ for any
constant $c>1$ \hspace{-2pt}
\\ \hline \multirow{4}{*}{\begin{tabular}[c]{@{}l@{}}
  \\ \hspace{-2pt}static\hspace{-20pt}\\
\hspace{-2pt}secure\hspace{-20pt}\\
\hspace{-2pt}sensor\hspace{-20pt}\\
\hspace{-2pt}networks\hspace{-20pt}\end{tabular}} &
\multirow{2}{*}{\hspace{-2pt}disk model\hspace{-20pt}} &
\hspace{-2pt}torus \hspace{-2pt}  &  \hspace{-2pt}connected if
$\frac{1}{q!}  \cdot \frac{{K_n}^{2q}}{{P_n} ^{q}}\cdot \pi{r_n}^{2}
\geq \frac{c\ln (n-m)}{n-m}$ for any constant $c>1$ \hspace{-2pt}
\\ \cline{3-4}
                                                                                    &                                                                                           &  \hspace{-2pt}square\hspace{-20pt} &  \hspace{-2pt}connected if $\frac{1}{q!}  \cdot \frac{{K_n}^{2q}}{{P_n} ^{q}} \cdot \pi{r_n}^{2} \geq \frac{c\ln (n-m)}{n-m}$ for any constant $c>c_n^{*}$ \hspace{-2pt}                                                                                                                         \\ \cline{2-4}
                                                                                    & \multirow{2}{*}{\begin{tabular}[c]{@{}l@{}} \hspace{-2pt}disk model w/\hspace{-20pt}\\  \hspace{-2pt}unreliable links\hspace{-20pt}\end{tabular}} &  \hspace{-2pt}torus\hspace{-20pt}  &  \hspace{-2pt}connected if $\frac{1}{q!}  \cdot \frac{{K_n}^{2q}}{{P_n} ^{q}}\cdot \pi{r_n}^{2} \cdot t_n \geq \frac{c\ln (n-m)}{n-m}$ for any constant $c>1$                                                                                                                      \\ \cline{3-4}
                                                                                    &                                                                                           &  \hspace{-2pt}square\hspace{-20pt} &  \hspace{-2pt}connected if $\frac{1}{q!}  \cdot \frac{{K_n}^{2q}}{{P_n} ^{q}} \cdot \pi{r_n}^{2} \cdot t_n \geq \frac{c\ln (n-m)}{n-m}$ for any constant $c>c_n^{\#}$                                                                                                              \\ \hline
\multirow{4}{*}{\begin{tabular}[c]{@{}l@{}}
\\ \hspace{-2pt}mobile\hspace{-20pt}\\
\hspace{-2pt}secure\hspace{-20pt}\\  \hspace{-2pt}sensor\\
\hspace{-2pt}networks\hspace{-20pt}\end{tabular}} &
\multirow{2}{*}{\hspace{-2pt}disk model\hspace{-20pt}} &
\hspace{-2pt}torus\hspace{-20pt}  &  \hspace{-2pt}connected for at
least $(n-m)^{c-1-\epsilon}$ consecutive time slots if $\frac{1}{q!}
\hspace{-2pt}
 \cdot  \hspace{-2pt}\frac{{K_n}^{2q}}{{P_n} ^{q}}\hspace{-2pt}\cdot\hspace{-2pt} \pi{r_n}^{2}\hspace{-2pt}
 \geq \hspace{-2pt}
\frac{c\ln (n-m)}{n-m}$ for any constant $c>1$                          \\
\cline{3-4}
                                                                                    &                                                                                           &  \hspace{-2pt}square\hspace{-20pt} &  \hspace{-2pt}connected for at least $(n-m)^{c-c_n^{*}-\epsilon}$ consecutive time slots if $\frac{1}{q!}  \hspace{-2pt} \cdot \hspace{-2pt} \frac{{K_n}^{2q}}{{P_n} ^{q}} \hspace{-2pt} \cdot \hspace{-2pt} \pi{r_n}^{2} \hspace{-2pt} \geq \hspace{-2pt} \frac{c\ln (n-m)}{n-m}$ for any constant $c>c_n^{*}$             \\ \cline{2-4}
                                                                                    & \multirow{2}{*}{\begin{tabular}[c]{@{}l@{}} \hspace{-2pt}disk model w/\hspace{-20pt}\\  \hspace{-2pt}unreliable links\hspace{-20pt}\end{tabular}} &  \hspace{-2pt}torus\hspace{-20pt}  &  \hspace{-2pt}connected for at least $(n-m)^{c-1-\epsilon}$ consecutive time slots if $\frac{1}{q!} \hspace{-2pt} \cdot \hspace{-2pt} \frac{{K_n}^{2q}}{{P_n} ^{q}} \hspace{-2pt} \cdot \hspace{-2pt} \pi{r_n}^{2} \hspace{-2pt} \cdot \hspace{-2pt} t_n \hspace{-2pt} \geq \hspace{-2pt} \frac{c\ln (n-m)}{n-m}$ for any constant $c>1$                \\ \cline{3-4}
                                                                                    &                                                                                           &  \hspace{-2pt}square\hspace{-20pt} &  \hspace{-2pt}connected for at least $(n-m)^{c-c_n^{\#}-\epsilon}$ consecutive time slots if $\frac{1}{q!} \hspace{-2pt} \cdot \hspace{-2pt} \frac{{K_n}^{2q}}{{P_n} ^{q}} \hspace{-2pt} \cdot \hspace{-2pt} \pi{r_n}^{2} \hspace{-2pt} \cdot \hspace{-2pt} t_n \hspace{-2pt} \geq \hspace{-2pt} \frac{c\ln (n-m)}{n-m}$ for any constant $c>c_n^{\#}$\hspace{-5pt} \\ \hline
\end{tabular}} \vspace{2pt}
  \caption{Different settings of secure sensor networks and their connectivity results
  under a random capture of $m$ nodes, where ``w/'' is short for ``with''. Note that $c_n^{*} $ is $ \max \Big\{1 +  \big(
\ln \frac{P_n}{{K_n}^2} \big) \big/ [\ln (n-m)] , ~4 \big( \ln
\frac{P_n}{{K_n}^2} \big)\big/ [\ln (n-m)]  \Big\} $, $c_n^{\#}$ is $\max \Big\{1 + \big( q \ln \frac{P_n}{{K_n}^2} + \ln \frac{1}{t_n} \big) \big/ [\ln (n-m)] ,
~4 \big( q \ln \frac{P_n}{{K_n}^2} + \ln \frac{1}{t_n} \big) \big/ [\ln (n-m)] \Big \} ${\color{white}, and $\epsilon$ is an arbitrary positive constant. All
results are in the asymptotic sense to hold with high probability,
and thus the conditions in the table only need to hold for all $n$
sufficiently large. Except the result under full visibility which
requires conditions on $K_n$ and $P_n$ in condition set $\Lambda$ in (\ref{Lambda-set}) on Page \pageref{Lambda-set}
but not the condition $r_n=o(1)$ in $\Lambda$, all other results
need the whole condition set $\Lambda$.}}
\label{tb-all-res-new}
\end{table*}

\subsection{{Unsplittability} of Secure Sensor Networks with the $q$-Composite Scheme}

Recall the definition of unsplittability in Section
\ref{motivation}. A secure sensor network is unsplittable if with
high probability an adversary that has captured an arbitrary set of
$o(n)$ sensors cannot partition the network into two chunks, which
both have linear sizes of sensors, and either are isolated from each
other or only have compromised communications in between.

\subsubsection{Unsplittability under Full Visibility}

\begin{thm} \label{thm-Unsplittability-full}

A secure sensor network with the $q$-composite scheme under full visibility is \emph{unsplittable} with high probability
if $\frac{1}{q!}  \cdot \frac{{K_n}^{2q}}{{P_n} ^{q}} \geq \frac{c\ln n}{n}$ for any constant $c>1$, $K_n = \omega( \ln n)$ and $K_n =
o\big(\min\{\sqrt{P_n}, \frac{P_n}{n}\}\big)$.

\end{thm}

By Theorem \ref{Connect-full-visi}, the conditions in Theorem
\ref{thm-Unsplittability-full} also ensure that a secure sensor
network with the $q$-composite scheme under full visibility is
connected with high probability. Intuitively, these conditions to guarantee connectivity in the absence of node capture can also ensure that when $o(n)$ nodes are compromised, almost all of the remaining nodes are still securely connected. More formally, the idea to show Theorem \ref{Connect-full-visi} is that even if $n_1 = o(n)$ nodes are compromised, we still have a securely connected network consisting of $n-n_2$ nodes, where $n_2$ can also be written as $o(n)$ despite $n_2 \geq n_1$. The reason for $n_2 \geq n_1$ is that capturing
nodes enables an adversary to compromise communications between
non-captured nodes which happen to use keys that are also shared by
captured nodes.

\begin{figure}[!t]
\begin{center}
  \includegraphics[scale=1]{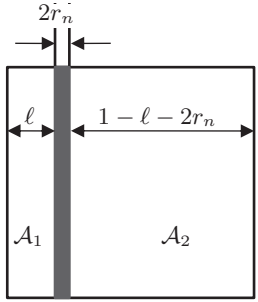}
  \caption{An adversary that has captured all $o(n)$ nodes in the shaded region can partition the network into two chunks $\mathcal
  {A}_1$ and $\mathcal
  {A}_2$, which both have
linear sizes of sensors with high probability and are isolated from
each other. We let $\ell$ be a constant with $0<\ell<1$, and consider $r_n = o(1)$.}   \label{fig-cap-square}
  \end{center}
  \end{figure}

\subsubsection{Unsplittability under the Disk Model}

\begin{thm} \label{thm-Unsplittability-disk}

A static secure sensor
network with the $q$-composite scheme under the disk model on the
unit torus or unit square is \emph{splittable} with high probability if $r_n = o(1)$.
%

%

\end{thm}

 To explain Theorem \ref{thm-Unsplittability-disk}, Figure \ref{fig-cap-square} illustrates that an adversary capturing all $o(n)$ nodes in the shaded region can partition the network into two chunks $\mathcal
  {A}_1$ and $\mathcal
  {A}_2$, which both have
linear sizes of sensors with high probability and are isolated from
each other. We present the specific details below for an intuitive understanding, and provide the technical proofs  in Section \ref{sec-prf-thm-Unsplittability-disk}. First, under  $r_n = o(1)$, we show that the number of nodes in the shaded region (i.e., a rectangle of  width $2r_n$ and
  length $1$) contains  $o(n)$ nodes with high probability  since the area of this region is also $o(1)$. Furthermore, because $\ell$ in Figure \ref{fig-cap-square} is a constant with $0<\ell<1$, the areas of chunks $\mathcal
  {A}_1$ and $\mathcal
  {A}_2$ are both constants, which enable us to prove that each chunk has  $\Theta(n)$ nodes (i.e., a constant fraction of the $n$ nodes) with high probability. Finally, it is clear that any node
in region $\mathcal
  {A}_1$ and any node
in region $\mathcal
  {A}_2$ have a distance at least $2r_n$, so the node set in $\mathcal
  {A}_1$ and the node set in $\mathcal
  {A}_2$ have no edge in between.

\subsection{{Unassailability} of Secure Sensor Networks with the $q$-Composite Scheme} \label{sec:resi}

Many prior studies have analyzed node-captured attacks, but they
often consider that the adversary \emph{randomly} captures sensors.
However, in practice, the adversary may tune its node-capture
strategy according to its goal. To analyze unassailability, we consider \textit{arbitrary} node-capture attacks, in which the adversary can
capture an {arbitrary} set of sensors.
Recalling the definition of unassailability  in Section \ref{motivation}, a secure sensor network is unassailable if
an
adversary that has captured an \textit{arbitrary} set of $o(n)$
sensors can compromise only $o(1)$ fraction of communication links
in the rest of the network; in other words, an adversary has to capture a
constant fraction of sensors to compromise a constant fraction of
communication links. For the $q$-composite scheme, previous studies have considered random node-capture attacks, but not {arbitrary} node-capture attacks formally. Hence, no prior work has rigorously
established the parameter conditions under which secure sensor
networks with the $q$-composite scheme are unassailable. Our Theorem \ref{thm:pcomp1} below presents the condition for unassailability of the $q$-composite scheme.

\begin{thm} \label{thm:pcomp1}
A secure sensor network with the $q$-composite scheme under any
communication model is unassailable with high probability under
${P_n}/{K_n} = \Omega(n)$.
\end{thm}

Under a different setting, a result similar to Theorem \ref{thm:pcomp1} is given in \cite{JZNodeCapture,zhao2015resilience}, but the result of \cite{JZNodeCapture,zhao2015resilience} is not for unassailability which by definition considers {arbitrary} node-capture attacks. Specifically, \cite{JZNodeCapture,zhao2015resilience} considers only \textit{random} node capture where the adversary captures a {random} set of $m$
sensors, whereas this paper addresses \textit{arbitrary} node capture where the adversary captures an {arbitrary} set of $m$
sensors. The choice
of $m$ nodes depends on the adversary's goal and also physical
limitations (e.g., some sensors may be easier to be captured than
others). Theorem \ref{thm:pcomp1} is proved in Section \ref{sec:lem:pcomp}.

 We check the practicality of $ {P_n}/{K_n} = \Omega(n)$ in Theorem \ref{thm:pcomp1}. Recall that the key ring size $K_n$ denotes the number of
keys in each sensor's memory and $P_n$ means the key pool size which the keys on each sensor are selected from. In
real-world sensor networks implementations, $K_n$ is several
orders of magnitude smaller than $n$ due to limited memory
and computational capability of sensors, and $P_n$ is larger
 than $n$ \cite{virgil,adrian,yagan}. Thus, condition ${P_n}/{K_n} =
\Omega(n)$
is practical.

To find the optimal $q$ that defends against node-capture attacks, we will investigate how $p_{\textnormal{compromised}}$ changes with respect to $q$. Since $p_{\textnormal{compromised}}$ depends on
the capture strategy, to study how $p_{\textnormal{compromised}}$ varies with respect to $q$, below we consider the special case where the adversary captures a \textit{random} set of $m$ sensors. This has been studied in \cite{JZNodeCapture}, but here we will improve the results of \cite{JZNodeCapture}. We state the research question more precisely as follows:
 given $n$, if we
fix the key ring size $K_n$ and the key-setup probability $p_q$, and
vary $q$, the required amount of key overlap, how does the
probability $p_{\textnormal{compromised}}$ change asymptotically
with respect to $q$? 

To answer the above question, we first  explain the key-setup probability $p_q$ below.
In the $q$-composite scheme, noting that all sensors' key rings are random variables following the same probability distribution, we define the key-setup probability $p_q$ as the
probability that the key rings of two sensors have at least $q$ keys
in common. More specifically, with $\{v_1,v_2,\ldots,v_n\}$ denoting the set of $n$ sensors in the network, and with $R_{\ell}$ denoting the key ring of sensor $v_{\ell}$ for $\ell=1,2,\ldots,n$, then we have the same value of $\mathbb{P}[ |R_i \cap R_j | \geq q]$ for each pair of different $i,j$ since each $R_{\ell}$ is constructed by selecting $K_n$ keys uniformly at random from the key pool $\mathcal{P}_n$ of $P_n$ keys. This value is denoted by $p_q$. Similarly, given $u$, we have the same value of $\mathbb{P}[ |R_i \cap R_j | =u]$ for each pair of different $i,j$ so we define $\rho_u$ as $\mathbb{P}[ |R_i \cap R_j | =u]$; i.e., $\rho_u$
is the probability that two nodes $v_i$ and $v_j$ share $u$ keys exactly in their key rings. Then we have $p_q = \mathbb{P}[ |R_i \cap R_j | \geq q] = \sum_{u=q}^{K_n}
  \mathbb{P}[|R_i \cap R_j | = u]  = \sum_{u=q}^{K_n} \rho_u$, where $\rho_u$ denoting $\mathbb{P}[|R_i \cap R_j | = u]$ can be computed as follows. After $R_i$ comprising $K_n$ keys is chosen from the key pool $\mathcal{P}_n$ of $P_n$ keys, to have $|R_{i} \cap R_{j}| = u$ with $R_j$ of $K_n$ keys choosing from $\mathcal{P}_n$, we need to select $u$ keys from $R_i$ and $(K_n-u)$ keys from $\mathcal{P}_n \setminus R_i$ to construct $R_j$, which gives $\binom{K_n}{u}\binom{P_n-K_n}{K_n-u}$ possibilities of $R_j$. Without  $|R_{i} \cap R_{j}| = u$, we have $\binom{P_n}{K_n}$ possible ways to find $K_n$ keys from $\mathcal{P}_n$ to form $R_j$. Given the above, we have
  \begin{align}
\rho_u = \mathbb{P}[|R_i \cap R_j | = u]&  =
\frac{\binom{K_n}{u}\binom{P_n-K_n}{K_n-u}}{\binom{P_n}{K_n}} \label{psq2sbIT}
\end{align}
and
\begin{align}
p_q = \mathbb{P}[ |R_i \cap R_j | \geq q] &  = \sum_{u=q}^{K_n}
\frac{\binom{K_n}{u}\binom{P_n-K_n}{K_n-u}}{\binom{P_n}{K_n}} . \label{psq2sbIT2}
\end{align}
 From (\ref{psq2sbIT2}), $p_q$ is a function of the parameters $K_n, P_n$ and $q$. Hence,
 to fix $p_q$, the key pool size $P_n$ needs to
decrease as $q$ increases.

We now present Theorem \ref{thm:pcomp2} below, which gives the optimal $q$ that defends against node-capture attacks.

\begin{thm}\label{thm:pcomp2}
In a secure sensor network with the $q$-composite scheme under any
communication model, if the key ring size $K_n$ and the key-setup
probability $p_q$ are both fixed, and conditions $K_n = \omega(1)$
and $K_n = o\big(\sqrt{P_n}\big)$ hold, after an adversary has
captured $m$ sensors with $m = o\Big(\sqrt{\frac{P_n}{{K_n}^2}}
\hspace{2pt}\Big)$, then the fraction of communications compromised
in the rest of the network asymptotically achieves its minimum
\begin{itemize}
  \item at $q = 1$ if $\frac{K_n}{m} < 2$,
  \item at
 both $q = 1$ and $q = 2$  if $\frac{K_n}{m} = 2$,
  \item at $q =
\big\lfloor{K_n}/{m}\big\rfloor$ if $\frac{K_n}{m} > 2$ and
${K_n}/{m}$ is not an integer, and
  \item at
 both $q = {K_n}/{m}$
and $q = ({K_n}/{m})- 1$ if ${K_n}/{m}$ is an integer greater than
$2$.
\end{itemize}

\end{thm}

Theorem \ref{thm:pcomp2} improves the result of \cite{JZNodeCapture}, and is proved in Section \ref{sec:thmpc2}.


\subsection{Design guidelines for secure sensor networks}

Based on our theorems above, we can provide   design guidelines of
secure sensor networks to achieve desired resilience and connectivity. We first take $\textnormal{static-network}^{\textnormal{disk
model}}_{\textnormal{unit torus}}$ as an example.

From Theorems \ref{static-torus} and \ref{thm:pcomp1}, for $\textnormal{static-network}^{\textnormal{disk
model}}_{\textnormal{unit torus}}$, to ensure i) the network is resilient in the sense that an adversary
capturing sensors at random has to obtain at least a constant
fraction of nodes of the network in order to compromise a constant
fraction of secure links, and ii) the network is connected with high probability, we can enforce ${P_n}/{K_n} = \Omega(n)$ and $\frac{1}{q!}
 \cdot \frac{{K_n}^{2q}}{{P_n} ^{q}}  \cdot \pi {r_n}^{2} \geq \frac{c\ln n}{n}$ with a constant $c>1$, under   condition set $\Lambda$ in (\ref{Lambda-set}) on Page \pageref{Lambda-set}. We recall that  $\Lambda$ includes $K_n= \omega(\ln n)$, $K_n= o\Big(\min\big\{\sqrt{P_n}, \frac{P_n}{n}\big\}\Big),$ and $ r_n = o(1)$. Given the above, we can set $K_n = c_1 (\ln n)^{1+\epsilon_1}$ for some $c_1$ and small $\epsilon_1$, $P_n = c_2 n (\ln n)^{1+\epsilon_2}$ for some $c_2$ and  $\epsilon_2>\epsilon_1$. Substituting these $K_n$ and $P_n$ to the above $\frac{1}{q!}
 \cdot \frac{{K_n}^{2q}}{{P_n} ^{q}}  \cdot \pi {r_n}^{2} \geq \frac{c\ln n}{n}$, we can set $r_n$ such that $\pi {r_n}^{2} \geq \frac{q!}{c} \cdot \frac{{c_2}^q}{{c_1}^{2q}} \cdot (\ln n)^{1-q(1+2\epsilon_1 - \epsilon_2)} n^{q-1}$.

 For $\textnormal{static-network}^{\begin{subarray}\textnormal{\textnormal{disk model}}\\ \textnormal{unit torus}\end{subarray}}_{\textnormal{unreliable links}}$, we can still use $K_n$ and $P_n$ above, while we set $r_n$ such that \mbox{$\pi {r_n}^{2} \cdot t_n \geq  \frac{q!}{c} \cdot \frac{{c_2}^q}{{c_1}^{2q}} \cdot (\ln n)^{1-q(1+2\epsilon_1 - \epsilon_2)} n^{q-1}$}, where $t_n$ is the probability of a link being active.

 For other networks, similarly we can use our theorems to present   guidelines for setting the network parameters.

\section{Useful Lemmas} \label{sec-preliminary-results}

\subsection{Coupling between random graphs}

We will couple different random graphs together. The idea is  converting a problem of one random graph to the corresponding problem in another random graph, in order to solve the original problem.
Formally, a coupling
\cite{zz,zhaoconnectivityTAC,2013arXiv1301.0466R,zhaoconnectivityTCNS,Krzywdzi} of two random graphs
$G_1$ and $G_2$ means a probability space on which random graphs
$G_1'$ and $G_2'$ are defined such that $G_1'$ and $G_2'$ have the
same distributions as $G_1$ and $G_2$, respectively. For notation
brevity, we simply say $G_1$ is a spanning subgraph\footnote{A graph
$G_a$ is a spanning subgraph (resp., spanning supergraph) of a graph
$G_b$ if $G_a$ and $G_b$ have the same node set, and the edge set of
$G_a$ is a subset (resp., superset) of the edge set of $G_b$.}
(resp., spanning supergraph) of $G_2$ if $G_1'$ is a spanning subgraph (resp., spanning supergraph) of $G_2'$.

 Using Rybarczyk's notation \cite{zz}, we
write
\begin{align}
G_1 \succeq & G_2 \quad (\textrm{resp.}, G_1 \succeq_{1-o(1)} G_2)
\label{g1g2coupling}
\end{align}
if there exists a coupling under which $G_2$ is a spanning subgraph
of $G_1$ with probability $1$ (resp., $1-o(1)$).

\begin{lem}[Rybarczyk \cite{zz}]  \label{mono-gcp}

For two random graphs $G_1$ and $G_2$, the following results hold
for any monotone increasing graph property $\mathcal {I}$.

\begin{itemize}
  \item If $G_1 \succeq G_2$, then $\mathbb{P} \big[ G_1\textrm{
has $\mathcal {I}$.}\big] \geq \mathbb{P} \big[ G_2\textrm{ has
$\mathcal {I}$.}\big]$.
  \item If $G_1 \succeq_{1-o(1)}  G_2$, then $$\mathbb{P} \big[ G_1\textrm{
has $\mathcal {I}$.}\big] \geq \mathbb{P} \big[ G_2\textrm{ has
$\mathcal {I}$.}\big]- o(1).$$
\end{itemize}

\end{lem}

\begin{lem}[Rybarczyk \cite{zz}]  \label{mono-gcp-cafpd}

For three random graphs $G_1$, $G_2$ and $G_3$, if $G_1
\succeq_{1-o(1)}  G_2$, then $G_1 \bcap G_3 \succeq_{1-o(1)}  G_2
\bcap G_3$.

\end{lem}

We use    $G_q(n, K_n,
P_n)$ to represent the graph model induced by the $q$-composite key predistribution scheme. In graph $G_q(n, K_n,
P_n)$ defined on $n$ nodes, each
node independently selects $K_n$ different keys {uniformly at random} from the same pool of $P_n$ distinct keys, and two nodes establish an edge in between if and only if they have at least $q$ keys
in common. Also, an Erd\H{o}s--R\'{e}nyi graph
\cite{citeulike:4012374} $G_{ER}(n, p_n)$ is defined on a set of $n$
nodes such that any two nodes establish an edge in between
independently with probability $p_n$.
 Lemma \ref{lem-cpgraph-rigrig} below relates graph $G_q(n, K_n,
P_n)$ with an Erd\H{o}s--R\'enyi graph.

\begin{lem} \label{lem-cpgraph-rigrig}
If $K_n = \omega\big(\max\{\ln n, \frac{\sqrt{P_n}}{n}\}\big)$ and
$K_n = o\big(\min\{\sqrt{P_n}, \frac{P_n}{n}\}\big)$, then
\begin{align}
  G_q(n,K_n,P_n) & \succeq_{1-o(1)}\textstyle{G_{ER}(n,\frac{1}{q!} \cdot \frac{{K_n}^{2q}}{{P_n} ^{q}} \cdot [1-o(1)] )}. \label{GerGurignbssd}
\end{align}

\end{lem}

 To prove Lemma \ref{lem-cpgraph-rigrig}, we
introduce an auxiliary graph called the \emph{binomial
$q$-intersection
 graph} $H_q(n,s_n,P_n)$ \cite{Rybarczyk,Assortativity,bloznelis2013}, which
 can be defined on $n$ nodes by the following process.
  There exists a key pool of size $P_n$. Each key in the
pool is added to each node \emph{independently} with probability
$s_n$. After each node obtains a set of keys, two nodes establish an edge in between if and only if they share at least $q$ keys. Clearly, the only difference between binomial $q$-intersection
 graph $H_q(n,s_n,P_n)$ and uniform $q$-intersection
 graph $G_q(n,K_n,P_n)$ is that in the former,
  the number of keys assigned to each
 node obeys a binomial distribution with $P_n$ as
the number of trials, and with $s_n$ as the success probability in
each trial, while in the latter graph, such number equals $K_n$ with
probability $1$.

In \cite{IEEE-TSIPN-17}, we prove Lemma \ref{lem-cpgraph-rigrig} by using
Lemmas \ref{brig_urig} and \ref{er_brig} below.

\begin{lem} \label{brig_urig}
If $K_n = \omega(\ln n)$ and $K_n = o\big(\sqrt{P_n}\big)$, with
$s_n$ set by
\begin{align}
 s_n   = \textstyle{\frac{K_n}{P_n}
 \Big(1 - \sqrt{\frac{3\ln
n}{K_n }}\hspace{2pt}\Big)}, \label{pnKn}
 \end{align}
then 
 it holds that
\begin{align}
  G_q(n,K_n,P_n) & \succeq_{1-o(1)}H_q(n,s_n,P_n). \label{eq_brig_urig}
\end{align}

\end{lem}

\begin{lem} \label{er_brig}
If $s_n  P_n = \omega(\ln n)$, $n{s_n}=o(1)$, $P_n{s_n}^2 = o(1)$
and $n^2{s_n}^2 P_n
 = \omega(1)$, then
there exits some $p_n$ satisfying
\begin{align}
p_n & = \textstyle{\frac{(P_n{s_n}^2)^q}{q!}} \cdot [1- o(1)]
\label{pnpb01}
\end{align}
such that Erd\H{o}s--R\'{e}nyi graph $G_{ER}(n,p_n)$
\cite{citeulike:4012374} obeys
\begin{align}
 H_q(n,s_n,P_n)& \succeq_{1-o(1)} G_{ER}(n,p_n) . \label{GerGb}
\end{align}

\end{lem}


\subsection{Useful results in the literature}

\begin{lem}[Erd\H{o}s and R\'{e}nyi \cite{citeulike:4012374}]  \label{erdsocon}

An Erd\H{o}s--R\'enyi graph $G_{ER}(n,p_n)$ is connected with high
probability if $p_n  \geq \frac{c\ln n}{n}$ for all $n$ sufficiently
large,
 where $c$ is a constant with $c>1$.

\end{lem}

The following two lemmas present topological properties (including connectivity) of graph
$\mathbb{G}(n, p_n , r_n, \mathcal{A}) $. Graph $\mathbb{G}(n, p_n , r_n, \mathcal{A}) $ is constructed as
follows. Let $n$ nodes be uniformly and independently deployed in
some network area $\mathcal{A}$, which is either a unit torus
$\mathcal{T}$ or a unit square $\mathcal{S}$. First, two nodes need
to have a distance of no greater than $r_n$ for having an edge in
between, which models the transmission constraints of nodes in
wireless networks. Then each edge between two nodes is preserved with
probability $p_n$ and deleted with
probability $1-p_n$, to model the link unreliability of wireless
links.
\begin{lem}[Penrose
\cite{Penrose2013}]  \label{lem-cpgraph-connew} The following
results hold.
\begin{itemize}
\item Under $r_n = o(1)$, graph $\mathbb{G}(n, p_n   , r_n, \mathcal{T}) $ is \emph{connected} with high probability
if $p_n  \cdot \pi {r_n}^{2} \geq \frac{c\ln n}{n}$ for all $n$
sufficiently large,
 where $c$ is a constant with $c>1$.
\item Under $r_n = o(1)$, graph $\mathbb{G}(n, p_n , r_n, \mathcal{S}) $ is \emph{connected} with high probability
if $p_n \cdot \pi {r_n}^{2} \geq \frac{c\ln n}{n}$ for all $n$
sufficiently large,
 where $c$ is a constant satisfying $c>  c_n^{*} $, with $c_n^{*}$ defined in (\ref{cnstar}).
\end{itemize}

\end{lem}

\begin{lem}[Penrose
\cite{Penrose2013}] \label{lem-lambda} ~
\begin{itemize}
\item The following results hold in
 graph $\mathbb{G}(n, p_n   , r_n, \mathcal{T}) $ under $r_n = o(1)$.
\begin{itemize}
\item The expected number of isolated nodes in graph $\mathbb{G}(n, p_n   , r_n, \mathcal{T}) $ is
asymptotically equivalent to $n e^{-\pi r_n^2 p_n n}$.
\item The probability of graph $\mathbb{G}(n, p_n   , r_n, \mathcal{T}) $
being disconnected is asymptotically equivalent to $1 -
e^{-\lambda_1}$, where $\lambda_1$ denotes the expected number of
isolated nodes in graph $\mathbb{G}(n, p_n   , r_n, \mathcal{T}) $.
\end{itemize}
\item The following results hold in
 graph $\mathbb{G}(n, p_n   , r_n, \mathcal{S}) $ under $r_n = o(1)$.
\begin{itemize}
\item The expected number of isolated nodes in graph
 $\mathbb{G}(n, p_n   , r_n, \mathcal{S}) $ is
asymptotically equivalent to $n e^{-\pi r_n^2 p_n n /
\widetilde{c_n}}$, where\\ $\widetilde{c_n} := \max \big\{1 +  \big(
\ln \frac{1}{p_n} \big) \big/ \ln n , ~4  \big( \ln \frac{1}{p_n}
\big) \big/ \ln n \big\} $.
\item The probability of graph $\mathbb{G}(n, p_n   , r_n, \mathcal{S}) $
being disconnected is asymptotically equivalent to $1 -
e^{-\lambda_2}$, where $\lambda_2$ denotes the expected number of
isolated nodes in graph
 $\mathbb{G}(n, p_n   , r_n, \mathcal{S}) $.
\end{itemize}
\end{itemize}
\end{lem}

\section{Establishing the Theorems} \label{prove:thm}

\subsection{Establishing Theorem
\ref{Connect-full-visi}} \label{secprf-Connect-full-visi}




We denote the graph topology of
$\textnormal{network}^{\textnormal{full visibility}}$ by $G_q(n,
K_n, P_n)$, which is known as a kind of random intersection graph
\cite{Rybarczyk,JZISIT14} in the literature. Graph $G_q(n, K_n,
P_n)$ is constructed on a node set with size $n$ as follows. Each
node is independently assigned a set of $K_n$ different keys,
selected {uniformly at random} from a pool of $P_n$ keys. An edge
exists between two nodes if and only if they have at least $q$ keys
in common.

From Lemma \ref{lem-cpgraph-rigrig}, there exists a coupling under
which $G_q(n, K_n, P_n)$ is an spanning supergraph of an
Erd\H{o}s--R\'{e}nyi graph $G_{ER}(n,\frac{1}{q!} \cdot
\frac{{K_n}^{2q}}{{P_n} ^{q}} \cdot [1-o(1)] )$ with high
probability, where an Erd\H{o}s--R\'{e}nyi graph
\cite{citeulike:4012374} $G_{ER}(n, p_n)$ is defined on a set of $n$
nodes such that any two nodes establish an edge in between
independently with probability $p_n$. Since connectivity is a
monotone increasing graph property\footnote{A graph property is
called monotone increasing if it holds under the addition of edges
in a graph.}, we obtain from Lemmas \ref{mono-gcp} and
\ref{lem-cpgraph-rigrig} that
\begin{align}
 & \mathbb{P} \big[ G_q(n, K_n, P_n)\textrm{
is connected.}\big] \nonumber \\ & \quad \geq    \mathbb{P} \big[
\textstyle{G_{ER}(n,\frac{1}{q!} \cdot \frac{{K_n}^{2q}}{{P_n} ^{q}}
\cdot [1-o(1)] )}\textrm{ is connected.}\big] - o(1). \label{bgqn}
 \end{align}

Given condition $\frac{1}{q!}
 \cdot \frac{{K_n}^{2q}}{{P_n} ^{q}}  \geq \frac{c\ln n}{n}$ for all $n$ sufficiently
 large
 with constant $c>1$, we use Lemma \ref{erdsocon} and have
\begin{align}
 & \mathbb{P} \big[
\textstyle{G_{ER}(n,\frac{1}{q!} \cdot \frac{{K_n}^{2q}}{{P_n} ^{q}}
\cdot [1-o(1)] )}\textrm{ is connected.}\big] \geq  1 - o(1),
 \end{align}
 which is substituted into (\ref{bgqn}) to derive
\begin{align}
 & \mathbb{P} \big[ G_q(n, K_n, P_n)\textrm{
is connected.}\big] \geq  1 - o(1).
 \end{align}
Hence, $G_q(n, K_n, P_n)$ is connected with high probability;
namely, $\textnormal{network}^{\textnormal{full visibility}}$  is
connected with high probability.

Although  the above results can also be derived using the results in \cite{ANALCO}, the proof techniques in  \cite{ANALCO} lack formality (specifically \cite[Equation (6.93)]{ANALCO}) and are different from the above proof. In addition, Bloznelis and Rybarczyk
\cite{Bloznelis201494} (resp., Bloznelis and {\L}uczak \cite{Perfectmatchings}) have recently investigated
  connectivity (resp., $k$-connectivity) of $\textnormal{network}^{\textnormal{full visibility}}$, but both results after a rewriting \vspace{1pt} address only the narrow and impractical range of $P_n =o\big(n^{\frac{1}{q}} (\ln n)^{-\frac{3}{5q}}\big)$. In contrast, from condition set $\Lambda$ in (\ref{Lambda-set}), our Theorem \ref{Connect-full-visi} considers a more practical range of $P_n = \omega(n \ln n)$. More details can be found in Section
\ref{sec:related-work}  for related work.

%
%
%

%

%

%

%
%

%
%
%


\subsection{Establishing Theorems \ref{static-torus}--\ref{static-square-unreliable}} \label{sec:static-torussquarepf}
%
%

%

To establish Theorems
\ref{static-torus}--\ref{static-square-unreliable}, we use an approach similar to that  for proving Theorem \ref{Connect-full-visi}. Specifically, we will show that the graph topology of each network in
Theorems \ref{static-torus}--\ref{static-square-unreliable} is an
\emph{spanning supergraph} of some graph $\mathcal {G}$ with high
probability, where graph $\mathcal {G}$ is connected with high
probability under the corresponding conditions. Since connectivity
is a monotone increasing graph property, we will complete proving
Theorems \ref{static-torus}--\ref{static-square-unreliable} by the
help of Lemma \ref{mono-gcp}.

For each network on an area
$\mathcal{A}$ in Theorems
\ref{static-torus}--\ref{static-square-unreliable} ($\mathcal{A}$ is
the unit torus $\mathcal{T}$ or the unit square $\mathcal{S}$), we will
show that its topology is an spanning supergraph of some graph
$\mathbb{G}(n, p_n , r_n, \mathcal{A}) $ with high probability.
Graph $\mathbb{G}(n, p_n , r_n, \mathcal{A}) $ is constructed as
follows. Let $n$ nodes be uniformly and independently deployed in
some network area $\mathcal{A}$, which is either a unit torus
$\mathcal{T}$ or a unit square $\mathcal{S}$. First, two nodes need
to have a distance of no greater than $r_n$ for having an edge in
between, which models the transmission constraints of nodes in
wireless networks. Then each edge between two nodes is preserved with
probability $p_n$ and deleted with
probability $1-p_n$, to model the link unreliability of wireless
links. 

The disk model induces a so-called random geometric
 graph $G_{RGG}(n, r_n, \mathbb{A})$, which is defined as follows. Let $n$ nodes
be uniformly and independently deployed in a network area
$\mathbb{A}$. An edge
  exists between two nodes if and only if their distance is no
   greater than $r_n$. With each link inactive with probability
   $t_n$,
    the link unreliability yields an Erd\H{o}s--R\'{e}nyi graph
 $G_{ER}(n, t_n)$. Then it is clear that graph $\mathbb{G}(n, p_n , r_n, \mathcal{A}) $
is the intersection\footnote{With two graphs $G_1$ and $G_2$ defined
on the same node set, two nodes have an edge in between in the
intersection $G_1\bcap G_2$ if and only if these two nodes have an
edge in $G_1$ and also have an edge in $G_2$.} of random geometric
 graph $G_{RGG}(n, r_n, \mathbb{A})$ and Erd\H{o}s--R\'{e}nyi graph
 $G_{ER}(n, p_n)$.

\begin{lem} \label{lem-cpgraph}
The following results hold under \\$K_n = \omega\big(\max\{\ln n,
\frac{\sqrt{P_n}}{n}\}\big)$ and $K_n = o\big(\min\{\sqrt{P_n},
\frac{P_n}{n}\}\big)$.
\begin{itemize}
\item The graph topology of $\textnormal{static-network}^{\textnormal{disk model}}_{\textnormal{unit torus}}$ is
a spanning supergraph of $\mathbb{G}(n, \frac{1}{q!}  \cdot \frac{{K_n}^{2q}}{{P_n} ^{q}} \cdot [1-o(1)]  , r_n, \mathcal{T}) $ with high probability.
\item The graph topology of $\textnormal{static-network}^{\begin{subarray}\textnormal{\textnormal{disk model}}\\ \textnormal{unit torus}\end{subarray}}_{\textnormal{unreliable links}}$ is
a spanning supergraph of \\$\mathbb{G}(n, \frac{1}{q!}  \cdot
\frac{{K_n}^{2q}}{{P_n} ^{q}} \cdot  t_n  \cdot [1-o(1)]   ,  r_n,
\mathcal{T}) $ with high probability.
\item The graph topology of $\textnormal{static-network}^{\textnormal{disk model}}_{\textnormal{unit square}}$ is
a spanning supergraph of $\mathbb{G}(n, \frac{1}{q!}  \cdot \frac{{K_n}^{2q}}{{P_n} ^{q}}   \cdot [1-o(1)]  , r_n, \mathcal{S}) $ with high probability.
\item The graph topology of $\textnormal{static-network}^{\begin{subarray}\textnormal{\textnormal{disk model}}\\ \textnormal{unit square}\end{subarray}}_{\textnormal{unreliable links}}$ is
a spanning supergraph of \\$\mathbb{G}(n, \frac{1}{q!}  \cdot
\frac{{K_n}^{2q}}{{P_n} ^{q}} \cdot  t_n  \cdot [1-o(1)]   ,r_n,
\mathcal{S}) $ with high probability.
\end{itemize}

\end{lem}

\subsubsection{Proving Theorems
\ref{static-torus}--\ref{static-square-unreliable} by Lemmas
\ref{lem-cpgraph-connew} and \ref{lem-cpgraph}}~

First, we use Lemma \ref{lem-cpgraph-connew} to obtain the
connectivity results of different graphs $\mathbb{G}(n, p_n , r_n,
\mathcal{A}) $ in Lemma \ref{lem-cpgraph}. Note that if $p_n  \cdot
\pi {r_n}^{2} \geq \frac{c\ln n}{n}$ for some constant $c>1$ (resp.,
$c>  c_n^{*} $), then $p_n  \cdot \pi {r_n}^{2}  \cdot [1-o(1)] \geq
\frac{c'\ln n}{n}$ for all $n$ sufficiently large, where $c'$ is any
constant satisfying $1<c'<c$ (resp., $c_n^{*}<c'< c$). Then from
Lemma \ref{lem-cpgraph-connew}, under $r_n = o(1)$, we have the
following results.
\begin{itemize}
\item Graph $\mathbb{G}(n, \frac{1}{q!}  \cdot \frac{{K_n}^{2q}}{{P_n} ^{q}} \cdot [1-o(1)]  , r_n, \mathcal{T}) $
is connected with high probability if $\frac{1}{q!}  \cdot
\frac{{K_n}^{2q}}{{P_n} ^{q}} \cdot \pi {r_n}^{2} \geq \frac{c\ln
n}{n}$ for any constant $c>1$.
\item Graph $\mathbb{G}(n, \frac{1}{q!}  \cdot \frac{{K_n}^{2q}}{{P_n} ^{q}} \cdot  t_n  \cdot [1-o(1)]   ,  r_n,
\mathcal{T}) $ is connected with high probability if $\frac{1}{q!}
\cdot \frac{{K_n}^{2q}}{{P_n} ^{q}} \cdot  t_n  \cdot \pi {r_n}^{2}
\geq \frac{c\ln n}{n}$ for any constant $c>1$.
\item Graph $\mathbb{G}(n, \frac{1}{q!}  \cdot \frac{{K_n}^{2q}}{{P_n} ^{q}} \cdot [1-o(1)]  , r_n, \mathcal{S}) $
is connected with high probability if $\frac{1}{q!}  \cdot
\frac{{K_n}^{2q}}{{P_n} ^{q}}  \cdot \pi {r_n}^{2} \geq \frac{c\ln
n}{n}$ for any constant $c>  c_n^{*} $.
\item Graph $\mathbb{G}(n, \frac{1}{q!}  \cdot \frac{{K_n}^{2q}}{{P_n} ^{q}} \cdot  t_n  \cdot [1-o(1)]   ,  r_n,
\mathcal{S}) $ is connected with high probability if $\frac{1}{q!}
\cdot \frac{{K_n}^{2q}}{{P_n} ^{q}}\cdot  t_n  \cdot \pi {r_n}^{2}
\geq \frac{c\ln n}{n}$ for any constant $c>  c_n^{\#} $.
\end{itemize}
Together with Lemma \ref{lem-cpgraph} (we will explain that
conditions in Lemma \ref{lem-cpgraph} all hold), the above results
complete proving Theorems
\ref{static-torus}--\ref{static-square-unreliable}, since any
spanning supergraph of a graph that is connected with high
probability is also connected with high probability.

We now show that conditions in Lemma \ref{lem-cpgraph} all hold.
First, condition set $\Lambda$ in (\ref{Lambda-set}) on Page \pageref{Lambda-set} hold in
Theorems \ref{static-torus}--\ref{static-square-unreliable}, so we
have $K_n = \omega(\ln n)$, $K_n = o\big(\min\{\sqrt{P_n},
\frac{P_n}{n}\}\big)$ and $r_n = o(1)$. Therefore, all conditions in
Lemma \ref{lem-cpgraph} will hold once we prove $K_n =
\omega\big(\frac{\sqrt{P_n}}{n}\big)$. Note that for each of
Theorems \ref{static-torus}--\ref{static-square-unreliable}, we
always have the condition $\frac{1}{q!}  \cdot
\frac{{K_n}^{2q}}{{P_n} ^{q}} \cdot \pi {r_n}^{2} = \omega\big(
\frac{\ln n}{n}\big)$, which along with $r_n = o(1)$ yields
$\frac{{K_n}^2}{P_n} = \omega\big(\big( \frac{\ln
n}{n}\big)^{1/q}\big) = \omega\big( \frac{1}{n^2}\big)$ and thus
$K_n = \omega\big(\frac{\sqrt{P_n}}{n}\big)$. \qeda

\subsubsection{Proof of Lemma
\ref{lem-cpgraph}}~

We consider the graph topologies of the networks in Lemma
\ref{lem-cpgraph} (i.e., those in Theorems
\ref{static-torus}--\ref{static-square-unreliable}).

First,  the $q$-composite key predistribution  scheme
is modeled by graph $G_q(n, K_n, P_n)$. The disk
model induces the so-called random geometric
 graph $G_{RGG}(n, r_n, \mathbb{A})$. Finally, in the presence of link
   unreliability, since each link is preserved with probability $t_n$, the
   underlying graph of link unreliability is an Erd\H{o}s--R\'{e}nyi graph
 $G_{ER}(n, t_n)$. Therefore, it is straightforward to see that the graph topologies of the
networks in Lemma \ref{lem-cpgraph} can be viewed as the
intersections of different random graphs.
Specifically, 
 we
obtain the following:
\begin{itemize}
  \item The graph topology of network $\textnormal{static-network}^{\textnormal{disk
model}}_{\textnormal{unit torus}}$ is $G_q(n, K_n, P_n)\cap
G_{RGG}(n , r_n, \mathcal{T}) $.
  \item The graph topology of network $\textnormal{static-network}^{\begin{subarray}\textnormal{\textnormal{disk
model}}\\ \textnormal{unit
torus}\end{subarray}}_{\textnormal{unreliable links}}$ is $G_q(n,
K_n, P_n)\cap G_{RGG}(n , r_n, \mathcal{T}) \cap G_{ER}(n , t_n) $.
  \item The graph topology of network $\textnormal{static-network}^{\textnormal{disk
model}}_{\textnormal{unit square}}$ is $G_q(n, K_n, P_n)\cap
G_{RGG}(n , r_n, \mathcal{S}) $.
  \item The graph topology of network $\textnormal{static-network}^{\begin{subarray}\textnormal{\textnormal{disk
model}}\\ \textnormal{unit
square}\end{subarray}}_{\textnormal{unreliable links}}$ is $G_q(n,
K_n, P_n)\cap G_{RGG}(n , r_n, \mathcal{S}) \cap G_{ER}(n , t_n) $.
\end{itemize}

From Lemmas \ref{mono-gcp-cafpd} and \ref{lem-cpgraph-rigrig},
%
 \f 
\begin{align}
& G_q(n, K_n, P_n)\cap G_{RGG}(n , r_n, \mathcal{T})
 \nonumber
\\  & \quad \succeq_{1-o(1)} \textstyle{\mathbb{G}(n, \frac{1}{q!}  \cdot \frac{{K_n}^{2q}}{{P_n} ^{q}} \cdot [1-o(1)]  , r_n, \mathcal{T})}.
\end{align}

Similarly, we have
\begin{align}
& G_q(n, K_n, P_n)\cap G_{RGG}(n , r_n, \mathcal{T}) \cap G_{ER}(n ,
t_n)
 \nonumber
\\  & \quad \succeq_{1-o(1)} \textstyle{\mathbb{G}(n, \frac{1}{q!}  \cdot \frac{{K_n}^{2q}}{{P_n} ^{q}} \cdot  t_n  \cdot [1-o(1)]   ,  r_n,
\mathcal{T})},
\\ & G_q(n, K_n, P_n)\cap
G_{RGG}(n , r_n, \mathcal{S})
 \nonumber
\\  & \quad \succeq_{1-o(1)} \textstyle{\mathbb{G}(n, \frac{1}{q!}  \cdot \frac{{K_n}^{2q}}{{P_n} ^{q}}   \cdot [1-o(1)]  , r_n, \mathcal{S})},
\end{align}
and
\begin{align}
 & G_q(n, K_n, P_n)\cap G_{RGG}(n , r_n, \mathcal{S}) \cap G_{ER}(n ,
t_n)
 \nonumber
\\  & \quad \succeq_{1-o(1)} \textstyle{\mathbb{G}(n, \frac{1}{q!} \cdot \frac{{K_n}^{2q}}{{P_n} ^{q}} \cdot  t_n  \cdot [1-o(1)]   , r_n, \mathcal{S}) }.
\end{align}

Then the proof of Lemma \ref{lem-cpgraph} is completed. \qeda


\subsection{Establishing Theorems
\ref{mobile-torus}--\ref{mobile-square-unreliable}}
\label{sec:mobile-torussquarepf}

On either the torus or the square, we consider i.i.d. mobility model.
Therefore, for all mobile networks that we consider, the overall
node distribution at each time slot is uniform, although the
position of each particular node may change over time. Then we can
view each mobile network at a single time slot as the corresponding
static network. Then we obtain the following Lemma \ref{mb-ntw-cp}
from Lemma \ref{lem-cpgraph}.

\begin{lem} \label{mb-ntw-cp}
The following results hold under $K_n = \omega\big(\max\{\ln n,
\frac{\sqrt{P_n}}{n}\}\big)$ and $K_n = o\big(\min\{\sqrt{P_n},
\frac{P_n}{n}\}\big)$.
\begin{itemize}
\item The graph topology of $\textnormal{mobile-network}^{\textnormal{disk model}}_{\textnormal{unit torus}}$ at any time slot $i$ is
a spanning supergraph of \\$\mathbb{G}(n, \frac{1}{q!}
 \cdot \frac{{K_n}^{2q}}{{P_n} ^{q}}\cdot [1-o(1)]  , r_n, \mathcal{T}) $
with high probability.
\item The graph topology of $\textnormal{mobile-network}^{\textnormal{disk model}}_{\textnormal{unit square}}$ at any time slot $i$ is
a spanning supergraph of \\$\mathbb{G}(n, \frac{1}{q!}
 \cdot \frac{{K_n}^{2q}}{{P_n} ^{q}}   \cdot [1-o(1)]  , r_n, \mathcal{S})
$ with high probability.
\item The graph topology of $\textnormal{mobile-network}^{\begin{subarray}\textnormal{\textnormal{disk model}}\\ \textnormal{unit torus}\end{subarray}}_{\textnormal{unreliable links}}$ at any time slot $i$ is
a spanning supergraph of \\$\mathbb{G}(n, \frac{1}{q!}
 \cdot \frac{{K_n}^{2q}}{{P_n} ^{q}} \cdot  t_n  \cdot [1-o(1)]   ,  r_n,
\mathcal{T}) $ with high probability.
\item The graph topology of $\textnormal{mobile-network}^{\begin{subarray}\textnormal{\textnormal{disk model}}\\ \textnormal{unit square}\end{subarray}}_{\textnormal{unreliable links}}$ at any time slot $i$ is
a spanning supergraph of \\$\mathbb{G}(n, \frac{1}{q!}
 \cdot \frac{{K_n}^{2q}}{{P_n} ^{q}} \cdot  t_n  \cdot [1-o(1)]   ,  r_n,
\mathcal{S}) $ with high probability.
\end{itemize}

\end{lem}

We now use Lemma \ref{mb-ntw-cp} to demonstrate Theorems
\ref{mobile-torus}--\ref{mobile-square-unreliable}. We first detail
the proof of Theorem \ref{mobile-torus}.

We define $\mathcal {C}_i$ as the event that network
$\textnormal{mobile-network}^{\textnormal{disk
model}}_{\textnormal{unit torus}}$ at time slot $i$ is connected.


Let $\lambda$ denote the expected number of isolated nodes in graph
$\mathbb{G}(n, \frac{1}{q!}  \cdot \frac{{K_n}^{2q}}{{P_n} ^{q}}
\cdot [1-o(1)]  , r_n, \mathcal{T}) $. Then from Lemma
\ref{lem-lambda} below, we have
\begin{align}
\lambda & \sim n e^{-\pi r_n^2 \cdot \frac{1}{q!}
 \cdot \frac{{K_n}^{2q}}{{P_n} ^{q}} \cdot n} . \label{lambda-val-fir}
\end{align}
Theorem \ref{mobile-torus} has the following condition:
\begin{align}
 \textstyle{\frac{1}{q!}  \cdot \frac{{K_n}^{2q}}{{P_n} ^{q}}  \cdot \pi
{r_n}^{2} \geq \frac{c\ln n}{n}}\textrm{ for some constant }c>1 ,
\end{align}
which yields
\begin{align}
n e^{-\pi r_n^2 \cdot \frac{1}{q!}
 \cdot \frac{{K_n}^{2q}}{{P_n} ^{q}} \cdot n} & \leq n e^{-c \ln n} = n^{1-c} .
\label{lambda-val-fir22}
\end{align}

From (\ref{lambda-val-fir}) (\ref{lambda-val-fir22}) and $c>1$, we
obtain
\begin{align}
\lambda &   \leq n^{1-c} \cdot [1+o(1)] , \label{lambda-val} 
\end{align}
and
\begin{align}
\lambda & = o(1).  \label{lambda-valo1}
\end{align}

From Lemma \ref{lem-lambda}, we also have
\begin{align}
 &  \mathbb{P}\Big[ \textnormal{$\textstyle{\mathbb{G}(n,
\frac{1}{q!}  \cdot \frac{{K_n}^{2q}}{{P_n} ^{q}} \cdot [1-o(1)]  ,
r_n, \mathcal{T})}$}
  \textrm{ is not connected.}
 \Big]  \nonumber
\\ & = \big(1 - e^{-\lambda}\big) \cdot [1\pm o(1)]  \nonumber
\\ & \leq 2\big(1 - e^{-\lambda}\big),\textnormal{ for all $n$ sufficiently large}.
\label{pttg}
\end{align}
Then from (\ref{pttg}) and Lemma \ref{mb-ntw-cp}, it follows for all
$n$ sufficiently large that
\begin{align}
 \mathbb{P}[\overline{\mathcal {C}_i}]
 & \leq 2\big(1 - e^{-\lambda}\big),\textnormal{ for all $n$ sufficiently
 large},\label{poct2}
\end{align}
yielding
\begin{align}
 \mathbb{P}[\mathcal {C}_i]
 & \geq 1 \hspace{-1pt}-  \hspace{-1pt}2\big(1  \hspace{-1pt}-  \hspace{-1pt}e^{-\lambda}\big)  \geq  1 \hspace{-1pt}- \hspace{-1pt}2\lambda,\textnormal{ for all $n$ sufficiently
 large} . \label{poct2ct}
\end{align}
where the last step uses the inequality $1 - e^{-x} \leq x$ for any
real $x$.



We now bound  the probability of event $\mathcal
{C}_1 \bcap \mathcal {C}_2 \bcap \cdots \bcap \mathcal {C}_{T}$
(i.e., the event that
$\textnormal{mobile-network}^{\textnormal{disk
model}}_{\textnormal{unit torus}}$ is connected for at least $T$
consecutive time slots from the beginning. Since we consider the i.i.d. mobility model, the events $\mathcal{C}_1, \ldots, \mathcal{C}_T$ are independent. Then we have
\begin{align}
& \mathbb{P}[\mathcal {C}_1  \bcap \mathcal {C}_2 \bcap \cdots \bcap
\mathcal {C}_{T}]  \nonumber
\\ & = \mathbb{P}[\mathcal {C}_1] \mathbb{P}[\mathcal {C}_2] \cdots
 \mathbb{P}[\mathcal {C}_T]
 \nonumber
\\ & \geq  (1-2\lambda)^{T}  \label{poct3gsgbx2bc},
\end{align}
where the last step uses (\ref{poct2ct}).

To evaluate $(1-2\lambda)^{T}$, we apply \cite[Fact
2]{ZhaoYaganGligor} which considers the Taylor series expansion and
have
\begin{align}
& \textstyle{(1-2\lambda)^{T} \geq 1 - 2\lambda T + \frac{1}{2}
(2\lambda)^2 T^2}  \label{poct3gsgbx}.
\end{align}
For $T= n^{c-1-\epsilon}$, we use (\ref{lambda-val}) to obtain
\begin{align}
\lambda T & \leq n^{1-c} \cdot [1+o(1)] \cdot n^{c-1-\epsilon} =
o(1)  ,
\end{align}
which is substituted into (\ref{poct3gsgbx}) to induce
\begin{align}
&  (1-2\lambda)^{T} \geq 1 - o(1). \label{poct3gsgbx2}
\end{align}

Applying (\ref{poct3gsgbx2}) to
(\ref{poct3gsgbx2bc}), we establish
\begin{align}
& \mathbb{P}[\mathcal {C}_1  \bcap \mathcal {C}_2 \bcap \cdots \bcap
\mathcal {C}_{T}] \to 1 - o(1).\label{poct3gsgbx2sfd}
\end{align}
In view of the fact that probability $\mathbb{P}[\mathcal {C}_1
\bcap \mathcal {C}_2 \bcap \cdots \bcap \mathcal {C}_{T}]$ is at
most $1$, we obtain from (\ref{poct3gsgbx2sfd}) that
\begin{align}
& \mathbb{P}[\mathcal {C}_1  \bcap \mathcal {C}_2 \bcap \cdots \bcap
\mathcal {C}_{T}] \to 1,\textrm{ as }n \to \infty.
\end{align}
Therefore, network $\textnormal{mobile-network}^{\textnormal{disk
model}}_{\textnormal{unit torus}}$ is connected for at least $T=
n^{c-1-\epsilon}$ consecutive time slots from the beginning.

The proofs of Theorems
\ref{mobile-torus-unreliable}--\ref{mobile-square-unreliable} using
Lemma \ref{lem-cpgraph} are similar to that of the above proof of
Theorem \ref{mobile-torus}. \qeda

\subsection{Proof of Theorem \ref{thm-Unsplittability-full}} \label{sec-prf-thm-Unsplittability-full}

We will prove Theorem \ref{thm-Unsplittability-full} by establishing
the following: after an adversary captures $m=o(n)$ nodes, with high
probability there exist $n-o(n)$ non-captured
    nodes which form
  a connected graph using only the non-compromised
  keys. In particular, we will show with high probability that there exist $L=n-o(n)$
non-captured
    nodes satisfying that \ding {172} each of these $L$ nodes has at least
$(1-\alpha)K_n$ keys that are not compromised by the adversary,
where $\alpha = 1- c^{-\frac{1}{6q}}$, and \ding {173} these $L$
nodes form
  a connected graph with each node using its $(1-\alpha)K_n$ non-compromised keys
  to discover neighbors.

We   use $\tau$
to denote the number of keys compromised by an adversary that has
captured $m=o(n)$ nodes. Clearly, $\tau \leq mK_n $ holds. Note that
we have $K_n = o\big(\frac{P_n}{n}\big)$ from condition $K_n =
o\big(\min\{\sqrt{P_n}, \frac{P_n}{n}\}\big)$. Let $X$ be the number
of a non-captured node's keys that are compromised by the adversary.
Since each node has $K_n$ keys selected from a $P_n$-size pool, the
expected value of $X$ is $K_n \cdot \tau/P_n  $.  From $\tau \leq
mK_n $, $K_n = o\big(\frac{P_n}{n}\big)$ and $m=o(n)$, \h
\begin{align}
\mathbb{E}[X] & = K_n \cdot \tau/P_n  = o(K_n).
\end{align}

We define $ \delta_n$ as the probability that a non-captured node
has at least $\alpha K_n$ keys that are compromised by the
adversary. From Markov's inequality, \f
\begin{align}
\delta_n   &    =\textstyle{  \mathbb{P}[X\geq \alpha K_n]
  \leq \frac{\mathbb{E}[X]}{\alpha K_n}  \leq
o(\alpha^{-1}) = o(1). }\label{deltao111}
\end{align}
%
%

Let $Y$ be the number of non-captured nodes, each of which has at
least $(1-\alpha)K_n$ keys that are not compromised by the
adversary. For two different non-captured nodes $x$ and $y$,
independence exists between how the key ring of node $x$ is
compromised by the adversary, and how the the key ring of node $y$
is compromised by the adversary. Then it is clear that $Y$ follows a
binomial distribution with parameters $n-m$ (the number of trials)
and $1-\delta_n$ (the success probability in each trial). From
Hoeffding's inequality, for any $\beta_n$, we have
\begin{align}
 \mathbb{P}[Y \leq (1-\delta_n - \beta_n)(n-m)]
 & \leq e^{-2{\beta_n}^2(n-m)} . \label{Yleq}
\end{align}
Setting $\beta_n = n^{-\frac{1}{3}}$, we obtain from (\ref{Yleq})
and $m=o(n)$ that
\begin{align}
 \mathbb{P}[Y \leq (1-\delta_n - n^{-\frac{1}{3}})(n-m)]
 & \leq e^{-2n^{-\frac{2}{3}}(n-m)} = o(1).
\end{align}
Therefore, with $L$ denoting $(1-\delta_n - n^{-\frac{1}{3}})(n-m)$,
we get
\begin{align}
 \mathbb{P}[Y \geq L]
 & \geq 1 - o(1)  .\label{pYM}
\end{align}
From (\ref{deltao111}) and $m=o(n)$, \h
\begin{align}
L = (1-\delta_n - n^{-\frac{1}{3}})(n-m) & = n-o(n). \label{pYM2}
\end{align}

Clearly, (\ref{pYM}) and (\ref{pYM2}) above prove the result \ding
{172} mentioned at the beginning of the proof; i.e., each of these
$L=n-o(n)$ nodes has at least $(1-\alpha)K_n$ keys that are not
compromised by the adversary.

 Now we show the result \ding {173}. We use Theorem \ref{Connect-full-visi}
 to show that the $L$ nodes form
  a connected graph with each node using its $(1-\alpha)K_n$ non-compromised keys
  to discover neighbors. From Theorem \ref{Connect-full-visi}, we
  need to have (i) $(1-\alpha)K_n = \omega(\ln n)$ and $ (1-\alpha)K_n =
   o\big(\min\big\{\sqrt{P_n}, \frac{P_n}{n}\big\}\big)$,
    which both clearly hold from the conditions of Theorem \ref{thm-Unsplittability-full}, and (ii) $\frac{1}{q!} \cdot \frac{{[(1-\alpha)K_n]}^{2q}}{{P_n} ^{q}} \geq
\widetilde{c}  \cdot \frac{\ln L}{L}$ for some constant
$\widetilde{c} >1$, which we prove below.

As mentioned, $\alpha$ equals $1- c^{-\frac{1}{6q}}$. We choose this
value of $\alpha$ since it leads to
\begin{align}
 \textstyle{\frac{1}{q!} \cdot \frac{{[(1-\alpha)K_n]}^{2q}}{{P_n} ^{q}}=
c^{-\frac{1}{3}}\cdot \frac{1}{q!}  \cdot \frac{{K_n}^{2q}}{{P_n}
^{q}}.} \label{pYM2st1}
\end{align}
From (\ref{pYM2}), we have $L\sim n$ so that $\frac{\ln L}{L} \sim
\frac{\ln n}{n}$, which along with $c>1$ yields for all $n$
sufficiently large that
\begin{align}
 \textstyle{\frac{\ln L}{L}
\leq c^{\frac{1}{3}} \cdot \frac{\ln n}{n}} . \label{lnLL}
\end{align}
Then from (\ref{pYM2st1}) (\ref{lnLL}) and condition $\frac{1}{q!}
\cdot \frac{{K_n}^{2q}}{{P_n} ^{q}} \geq \frac{c\ln n}{n}$, we
obtain for all $n$ sufficiently large that
\begin{align}
 \textstyle{\frac{1}{q!} \cdot \frac{{[(1-\alpha)K_n]}^{2q}}{{P_n}
 ^{q}}} \geq
c^{-\frac{1}{3}}\cdot \textstyle{c \frac{\ln n}{n} \geq
c^{\frac{1}{3}} \cdot \frac{\ln L}{L}}.
\end{align}
Thus we can choose constant $\widetilde{c}= c^{\frac{1}{3}} >1$ to
have $\frac{1}{q!} \cdot \frac{{[(1-\alpha)K_n]}^{2q}}{{P_n} ^{q}}
\geq \widetilde{c}  \cdot \frac{\ln L}{L}$ for all $n$ sufficiently
large. Then as explained above, result \ding {173} is proved from
Theorem \ref{Connect-full-visi}.

In view of results \ding {172} and \ding {173}, after an adversary
captures $m=o(n)$ nodes, with high probability, there exist
  $n-o(n)$ non-captured
    nodes, which form
  a connected graph using only the non-compromised
  keys. Therefore, with high probability, an adversary that has captured an arbitrary set of $o(n)$ sensors
{cannot} partition the network into two chunks, which both have
linear sizes of sensors, and either are isolated from each other or
only have compromised communications in between. Then the studied
secure sensor network with the $q$-composite scheme under full
visibility is unsplittable. \qeda

\subsection{Proof of Theorem \ref{thm-Unsplittability-disk}} \label{sec-prf-thm-Unsplittability-disk}

We show that under the disk model, an adversary that has captured some set of $o(n)$
sensors can partition the network into two chunks, which both have
linear sizes of sensors with high probability and are isolated from
each other.

\begin{figure}[!t]
\begin{center}
  \includegraphics[scale=1]{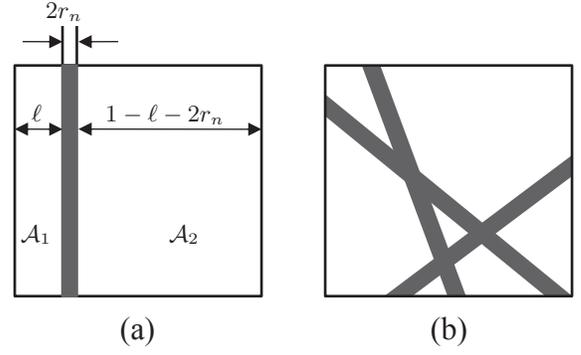}
  \caption{An adversary captures all nodes in the shaded region.} \label{splittability-eps}
  \end{center}
  \end{figure}

As shown in Figure \ref{splittability-eps}-(a) (Figure
\ref{splittability-eps}-(b) is explained later), the adversary
captures all nodes in the shaded region $\mathcal
  {A}_0$, which has a width $2r_n$ and
a length $1$. The region $\mathcal
  {A}_1$ on the left side of the shaded region has a
width $\ell$ and a length $1$, while the region $\mathcal
  {A}_2$ on the right side of
the shaded area has a width $1-\ell-2r_n$ and a length $1$, where
$\ell$ is a constant with $0<\ell<1$.

With $|\mathcal
  {A}_i|$ denoting the area of $ \mathcal
  {A}_i$ for each $i=0,1,2$, we have $|\mathcal
  {A}_0| = 2r_n = o(1)$, $|\mathcal
  {A}_1| = \ell = \Theta(1)$ and $|\mathcal
  {A}_2| = 1-\ell-2r_n = \Theta(1)$. Therefore, to show the number of nodes in $\mathcal
  {A}_0$ is $o(n)$ with high probability, and
   the numbers of nodes in $\mathcal
  {A}_1$ and $\mathcal
  {A}_2$ are both $\Theta(n)$ with high probability, it suffices to
  prove the following: for some region $\widetilde{\mathcal
  {A}}$ as a sub-field of the whole network region $ \mathcal
  {A}$, with $|\widetilde{\mathcal
  {A}}|$ denoting the area of $\widetilde{\mathcal
  {A}}$ and with $\widetilde{N}$ denoting the number of nodes in region $\widetilde{\mathcal
  {A}}$, we have \ding {182} if $|\widetilde{\mathcal
  {A}}| = o(1)$, then $\widetilde{N} = o(n)$ holds with high probability, and \ding {183} if $|\widetilde{\mathcal
  {A}}| = \Theta(1)$, then $\widetilde{N} = \Theta(n)$ holds with high probability.


   Since all $n$ nodes are uniformly distributed in the whole network region $\mathcal
  {A}$ with area $1$ ($\mathcal
  {A}$ is a unit torus or a unit square), 
 then $\widetilde{N}$ (the number of nodes in the region $\widetilde{\mathcal
  {A}}$) follows a binomial distribution with parameter $n$ as the
number of trials, and parameter $|\widetilde{\mathcal
  {A}}|$ as the success probability in
each trial. For any $\nu_n>0$, we use Hoeffding's inequality to have
\begin{align}
 &\mathbb{P}[(|\widetilde{\mathcal
  {A}}|-\nu_n)n \leq \widetilde{N} \leq (|\widetilde{\mathcal
  {A}}|+\nu_n)n]  \geq
 1- 2e^{-2{\nu_n}^2n} .
\end{align}

Setting $\nu_n = n^{-\frac{1}{3}}$, we have $2e^{-2{\nu_n}^2n} =
o(1)$. Then
\begin{align}
 &\mathbb{P}[(|\widetilde{\mathcal
  {A}}|-\nu_n)n \leq \widetilde{N} \leq (|\widetilde{\mathcal
  {A}}|+\nu_n)n] \geq 1-
 o(1).
\end{align}
Then we have
\begin{itemize}
  \item if $|\widetilde{\mathcal
  {A}}| = o(1)$, then $\widetilde{N} \leq (|\widetilde{\mathcal
  {A}}|+\nu_n)n = o(n)$ holds with high probability.
  \item if $|\widetilde{\mathcal
  {A}}| = \Theta(1)$, it holds with high probability that
   $(|\widetilde{\mathcal
  {A}}|-\nu_n)n \leq \widetilde{N} \leq (|\widetilde{\mathcal
  {A}}|+\nu_n)n $. From $|\widetilde{\mathcal
  {A}}| = \Theta(1)$ and $\nu_n = n^{-\frac{1}{3}}$, both $(|\widetilde{\mathcal
  {A}}|-\nu_n)n$ and $(|\widetilde{\mathcal
  {A}}|+\nu_n)n$ are $\Theta(n)$. Therefore, $\widetilde{N} =
  \Theta(n)$ follows with high probability.
\end{itemize}
Therefore, the results \ding {182} and \ding {183} above are proved.
Returning to Figure \ref{splittability-eps}-(a), as explained
before, in view of $|\mathcal
  {A}_0|  = o(1)$, $|\mathcal
  {A}_1|   = \Theta(1)$ and $|\mathcal
  {A}_2| = \Theta(1)$, we obtain the following: with $N_i$
   denoting the number of nodes in region $ \mathcal
  {A}_i$ for each $i=0,1,2$, we have $N_0 = o(n)$ with high probability, $N_1 = \Theta(n)$ with high probability, and
  $N_2 = \Theta(n)$ with high probability. Then $\big\{N_0 = o(n)\big\} \bcap  \big\{N_1 =
\Theta(n)\big\} \bcap \big\{N_2 = \Theta(n)\big\}$ occurs with
probability $1-o(1)$ by the union
  bound. Hence, $\big\{N_0 = o(n)\big\} \bcap  \big\{N_1 = \Theta(n)\big\}
\bcap \big\{N_2 = \Theta(n)\big\}$ holds with high probability.

%
%

From Figure \ref{splittability-eps}-(a), it is clear that any node
in region $\mathcal
  {A}_1$ and any node
in region $\mathcal
  {A}_2$ have a distance at least $2r_n$, so the node set in region $\mathcal
  {A}_1$ and the node set in region $\mathcal
  {A}_2$ have no edge in between. Hence, with high probability,
  an adversary that has captured $o(n)$ nodes in region $N_0$
 partitions the network into two chunks, which both have linear sizes
of sensors and are isolated from each other. Then the studied static
secure sensor network with the $q$-composite scheme under the disk
model on the unit torus or unit square is splittable. \qeda

\emph{Remarks:}
\begin{itemize}
  \item The adversary may split the network into multiple chunks instead of just two, as
  illustrated in Figure \ref{splittability-eps}-(b).
  \item A future direction is to extended the result to more general topology (here
  we consider a torus or square).
\end{itemize}

\subsection{Proof of Theorem \ref{thm:pcomp1}} \label{sec:lem:pcomp}

We will prove Theorem \ref{thm:pcomp1} by showing that under
${P_n}/{K_n} = \Omega(n)$, an adversary that has captured an
\textit{arbitrary} set of $o(n)$ sensors can only compromise $o(1)$ fraction
of communication links in the rest of the network.

Suppose the adversary has captured some set of $m$ nodes. The choice
of $m$ nodes depends on the adversary's goal and also physical
limitations (e.g., some sensors may be easier to be captured than
others). Let $E_{\tau}$ be the event that the $m$ captured nodes
have $\tau$ different keys in their key rings in total. It is clear
that $K_n \leq \tau \leq \min\{mK_n, P_n\}$. Recall that
$p_{\textnormal{compromised}}$ is the fraction of compromised
communications among non-captured nodes
($p_{\textnormal{compromised}}$ is also the probability that the
secure link between two non-captured nodes is compromised).
Conditioning on event $E_{\tau}$, let
$p_{\textnormal{compromised}}(\tau)$ be the fraction of compromised
communications among non-captured nodes (conditioning on event
$E_{\tau}$, $p_{\textnormal{compromised}}(\tau)$ is also the
probability that the secure link between two non-captured nodes is
compromised).
 By the law of total probability, we have
\begin{align}
p_{\textnormal{compromised}} & = \textstyle{
\sum_{\tau=K_n}^{\min\{mK_n, P_n\}} \big\{ \mathbb{P}[E_{\tau}]
\cdot p_{\textnormal{compromised}}(\tau)  \big\}  }. \label{problw}
\end{align}

The adversary's node capture strategy determines
$\mathbb{P}[E_{\tau}] $. With $K_n \leq \tau \leq \min\{mK_n,
P_n\}$, it holds that $\sum_{K_n}^{\min\{mK_n,
P_n\}}\mathbb{P}[E_{\tau}] = 1 $. Clearly, as $\tau$ increases,
$p_{\textnormal{compromised}}(\tau)$ increases (of course, if $\tau$
reaches $P_n$, $p_{\textnormal{compromised}}(\tau)$ becomes $1$ and
can not increase anymore). With $K_n \leq \tau \leq \min\{mK_n,
P_n\}$, then $p_{\textnormal{compromised}}(\tau)$ achieves its
maximum when $\tau =\min\{mK_n, P_n\}$. Therefore, for an adversary
that has captured \emph{any} set of $m$ sensors, it is simple to
derive from (\ref{problw}) that
\begin{align}
p_{\textnormal{compromised}} & \leq
p_{\textnormal{compromised}}(\min\{mK_n, P_n\}).
\label{p-compromised}
\end{align}

Under ${P_n}/{K_n} = \Omega(n)$ and $m=o(n)$, it holds that
\begin{align}
mK_n = o(P_n). \label{mKnPno}
\end{align}
  Then we have $mK_n \leq P_n$ and thus $\min\{mK_n, P_n\} = mK_n$   for all $n$
sufficiently large, which is used in (\ref{p-compromised}) to derive
\begin{align}
p_{\textnormal{compromised}} & \leq
p_{\textnormal{compromised}}(mK_n). \label{problw-newbspkkk2}
\end{align}
In addition, to
maximize $p_{\textnormal{compromised}}$ as
$p_{\textnormal{compromised}}(mK_n)$, the adversary can
capture some set of $m$ sensors which have $mK_n$
different keys in their key rings in total. Therefore, given (\ref{problw-newbspkkk2}), the proof of Theorem \ref{thm:pcomp1} is completed once
we demonstrate $p_{\textnormal{compromised}}(mK_n) = o(1)$.

By Yum and Lee
\cite{6170857}, it follows that
\begin{align}
\textstyle{p_{\textnormal{compromised}}(\tau)   =
\sum_{u=q}^{K_n}\bigg[ \frac{\binom{\tau}{u}}{\binom{P_n}{u}} \cdot
\frac{\rho_u}{p_{q}}\bigg]} ,\label{p-compromised-tau}
\end{align}
where the expressions of $\rho_u$ and $p_q$ are given in (\ref{psq2sbIT}) and (\ref{psq2sbIT2}).

To compute $p_{\textnormal{compromised}}(mK_n)$, we look at
$p_{\textnormal{compromised}}(\tau)$ given in
(\ref{p-compromised-tau}). Then we bound the term
$\binom{\tau}{u}\Big/\binom{P_n}{u}$ in (\ref{p-compromised-tau}).
With $q \leq u\leq K_n$, we obtain
\begin{align}
 \textstyle{\binom{\tau}{u}\big/\binom{P_n}{u}
 \leq \frac{\tau^u/(u!)}{{(P_n-u)}^u/(u!)}
 \leq \frac{\tau^q}{{(P_n-K_n)}^q}},
\end{align}
which along with $\sum_{u=q}^{K_n}\rho_u=p_{q}$ is used in
(\ref{p-compromised-tau}) to yield
\begin{align}
  & p_{\textnormal{compromised}}(\tau) \leq\textstyle{ \sum_{u=q}^{K_n}\big[
\frac{\tau^q}{{(P_n-K_n)}^q} \cdot \frac{\rho_u}{p_{q}}\big]
 = \frac{\tau^q}{{(P_n-K_n)}^q}}. \label{sumuq}
\end{align}
Setting $\tau$ as $mK_n$, we have
\begin{align}
p_{\textnormal{compromised}}(mK_n) \leq
\textstyle{\frac{(mK_n)^q}{{(P_n-K_n)}^q}} \label{pcmp_up}  = o(1),
\end{align}
where the last step uses $mK_n = o(P_n-K_n)$, which holds from
(\ref{mKnPno}). \qeda


\subsection{Proof of Theorem \ref{thm:pcomp2}} \label{sec:thmpc2}



Note that different from the case of Theorem \ref{thm:pcomp1}, in Theorem \ref{thm:pcomp2} here, we do not have $mK_n \leq P_n$ for all $n$
sufficiently large. Hence, we can have (\ref{p-compromised}) only, instead of having (\ref{problw-newbspkkk2}).

 When the adversary captures $m$ nodes \emph{randomly},
$\mathbb{P}[E_{\tau}]$ is \mbox{non-zero}  for all $\tau$ satisfying $K_n
\leq \tau \leq \min\{mK_n, P_n\}$. Under this \emph{random} attack,
the computation of $p_{\textnormal{compromised}}$ is not that
straightforward. Yet, \cite{JZNodeCapture} presents an asymptotic expression of $p_{\textnormal{compromised}}$ as Lemma \ref{lem:pcomp2} below.
\begin{lem} \label{lem:pcomp2} 
  If the adversary captures $m$ nodes \emph{randomly} with
  $m = o\big(\sqrt{\frac{P_n}{{K_n}^2}} \hspace{2pt}\big)$, then
$p_{\textnormal{compromised}} $ is asymptotically equivalent to
$\big(\frac{mK_n}{P_n} \big)^q$ under $K_n = \omega(1)$ and $K_n =
o\big(\sqrt{P_n}\big)$.

\end{lem}


From Lemma \ref{lem:pcomp2}, \h
\begin{align}
  p_{\textnormal{compromised}}   & \sim
\textstyle{\big(\frac{mK_n}{P_n} \big)^q}. \label{ptnq}
\end{align}

Under $K_n = \omega(1)$ and $K_n = o\big(\sqrt{P_n}\big)$, as given
in (\ref{pssimq}), we have $p_q  \sim \frac{1}{q!}
 \cdot \frac{{K_n}^{2q}}{{P_n} ^{q}} $, which with (\ref{ptnq}) above
induces
\begin{align}
 \textstyle{\frac{p_{\textnormal{compromised}}}{p_q}  \sim \big(\frac{mK_n}{P_n} \big)^q
  \cdot q! \cdot \frac{{P_n} ^{q}}{{K_n}^{2q}}  = q!
  \big(\frac{m}{K_n}\big)^q = f(q)},
\end{align}
where $f(q)$ is defined by
\begin{align}
\textstyle{f(q)  =  q! \big(\frac{m}{K_n}\big)^q}. \label{fq_expr}
\end{align}
We now analyze how $f(q)$ changes as $q$ varies, to understand the
asymptotic behavior of $p_{\textnormal{compromised}}$. From
(\ref{fq_expr}), we compute $f(q+1)/f(q)$ through
\begin{align}
\textstyle{\frac{f(q+1)}{f(q)}  = \frac{m(q+1)}{K_n}}. \label{fq1fq}
\end{align}
Then using (\ref{fq1fq}) to analyze the monotonicity of $f(q)$, we finally establish Theorem \ref{thm:pcomp2}.

\section{Simulation Results} \label{sec:expe}

We now provide simulation results to confirm the theoretical findings.
All plots in the paper are obtained from simulation. Specifically,  in each plot, the data points are obtained by averaging over
 $500$ independent network samples.

\subsection{Simulation Results for Connectivity}
%

\begin{figure}[!t]
\begin{center}
  \includegraphics[scale=0.5]{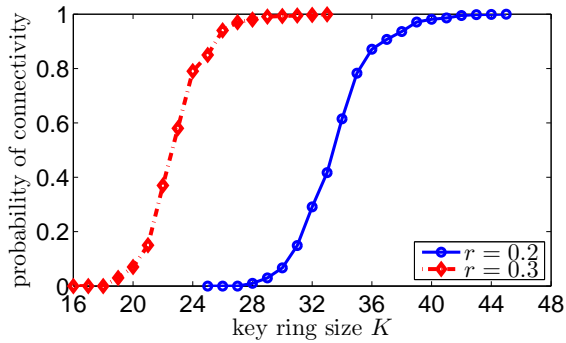}
  \caption{A plot of the empirical probability that network $\textnormal{static-network}^{\textnormal{disk model}}_{\textnormal{unit torous}}$
is connected as a function of the key ring size $K$ with $q=2$,
$n=2000$, $P=5000$, and $r=0.2,
0.3$.} \label{ccs-con1}
  \end{center}
  \end{figure}

\begin{figure}[!t]
\begin{center}
  \includegraphics[scale=0.5]{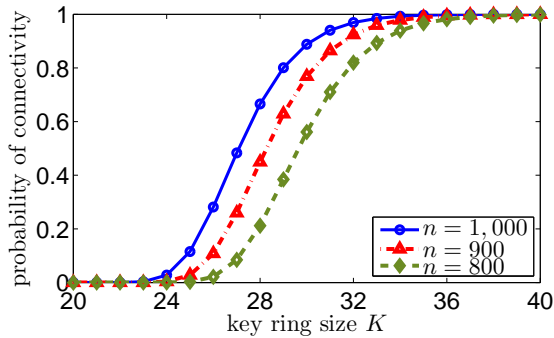}
  \caption{A plot of the empirical probability that network  $\bEaa$ after the random capture of 10 nodes
is connected as a function of the key ring size $K$ with $q=2$,
$n=1000, 900, 800$, $P=5000$, $r=0.3$ and $t=0.9$.}
\label{ccs-con2}
  \end{center}
  \end{figure}

We present below simulation results of connectivity in secure sensor
networks to confirm the theorems. Figure \ref{ccs-con1} depicts the
probability that network
$\textnormal{static-network}^{\textnormal{disk
model}}_{\textnormal{unit
  torous}}$ is
connected with $q=2$. We vary the key ring size $K$, and set
$n=2000$, $P=5000$, and $r=0.2,0.3$. For each pair $(K,r)$, we
generate $500$ independent samples of $\mathbb{G}_q(n, K, P,  r,
\mathcal {T})$ and record the count that the obtained graph has
connectivity. Then we derive the empirical probabilities after
dividing the counts by $500$. Similarly, in Figure \ref{ccs-con2},
we plot the probability that  network
$\bEaa$ is connected after an adversary captures 10 random
nodes, for $q=2$,
$n=1000, 900, 800$, $P=5000$, $r=0.3$ and $t=0.9$.
As illustrated in both figures, we observe the transitional
behavior of connectivity. Moreover, for both Figure \ref{ccs-con1} and \ref{ccs-con2}, substituting the parameter values into the corresponding theorems of this paper, we can confirm the correctness of the theorems.

\begin{figure}[!t]
\begin{center}
  \includegraphics[scale=0.5]{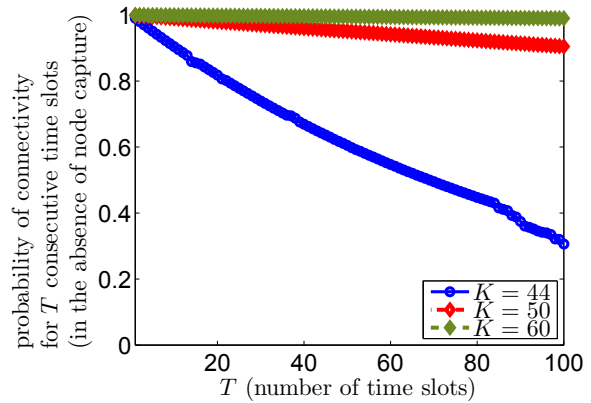}
  \caption{A plot of the empirical probability that mobile network $\textnormal{mobile-network}^{\textnormal{disk model}}_{\textnormal{unit
  square}}$
is connected for at least $ T$ consecutive time slots as a function
of the key ring size with $q=2$, $n=1000$, $P=6000$, $r=0.25$ and
$K=44,50,60$.}  \label{mobility-eps}
  \end{center}
  \end{figure}

Figure \ref{mobility-eps} presents a plot of the empirical
probability that mobile network
$\textnormal{mobile-network}^{\textnormal{disk
model}}_{\textnormal{unit
  square}}$
is connected for at least $ T$ consecutive time slots as a function
of the key ring size with $q=2$, $n=1000$, $P=6000$, $r=0.25$ and
$K=44,50,60$. We consider the i.i.d. mobility model, where each node independently picks a new location uniformly at random at the beginning of a time slot and stays at the location in the rest of the time slot. At each time slot, we check whether the
network is connected. We generate $500$ independent samples and
record the count that $\textnormal{mobile-network}^{\textnormal{disk
model}}_{\textnormal{unit
  square}}$ is connected for at least $ T$ consecutive time slots. Then we derive the empirical
probabilities after dividing the counts by $500$. As illustrated in
Figure \ref{mobility-eps}, incrementing $K$ increases the
probability that mobile network
$\textnormal{mobile-network}^{\textnormal{disk
model}}_{\textnormal{unit
  square}}$
is connected for at least $ T$ consecutive time slots.

\subsection{Simulation Results for Unassailability}

We provide simulation plots below for a comparison with the
theoretical results above.

\subsubsection{Varying $q$ with key ring size and key-setup prob. fixed}

We first fix the key ring size and the key-setup probability, and
vary $q$ to see how the probability $p_{\textnormal{compromised}}$
of link compromise changes.

\begin{figure}[!t]
\begin{center}
  \includegraphics[scale=0.5]{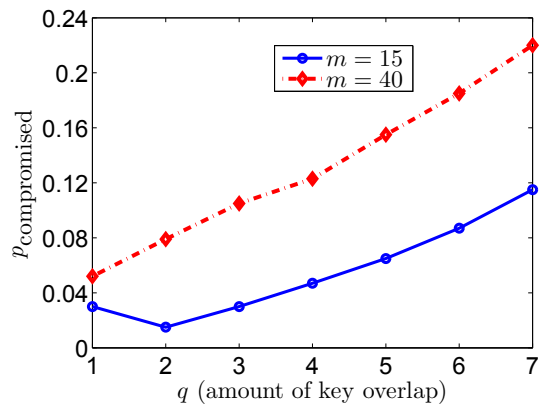}
  \caption{A plot of probability $p_{\textnormal{compromised}}$ of link compromise with
respect to $q$, under $K = 40$, $p_q = 0.1$ and $m = 15, 40$. To
ensure a fixed $p_q$, the key pool size $P$ needs to decrease as $q$
increases.}   \label{ccs-res}
  \end{center}
  \end{figure}

 Figure \ref{ccs-res} depicts probability
$p_{\textnormal{compromised}}$ with respect to $q$. We fix the key
ring size at $K =40$, and the key-setup probability at $p_q = 0.1$,
and set the number $m$ of captured nodes as $15$ or $40$. As
illustrated, for $m=15$, as $q$ increases,
$p_{\textnormal{compromised}}$ first decreases and then increases;
and for $m=40$, as $q$ increases, $p_{\textnormal{compromised}}$
always increases. These are in agreement with the analytical results. In particular, for $K =40$ and $m=15 $, given $ \frac{K}{m}  =
 \frac{40}{15} \approx 2.67$ and Theorem \ref{thm:pcomp2},  $p_{\textnormal{compromised}}$ takes its minimum at
$q = 2$, in accordance with the
fact that in Figure \ref{ccs-res}, $p_{\textnormal{compromised}}$
for $m=15$ achieves its minimum at $q=2$.  For $K =40$ and $m=40 $, both both   theoretical and experimental results show that  $p_{\textnormal{compromised}}$
takes the minimum at $q=2$.

\begin{figure}[!t]
\begin{center}
 \includegraphics[scale=0.5]{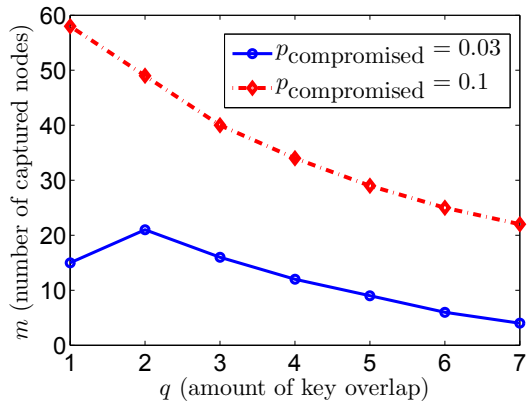}
  \caption{A plot under varying $q$ for
the number $m$ of nodes that the adversary has to capture to
compromise a link between two non-captured nodes with probability
$p_{\textnormal{compromised}}$. The parameters are $K = 40$, $p_q =
0.1$ and $p_{\textnormal{compromised}} = 0.03, 0.1$. To fix $p_q$,
the key pool size $P$ decreases as $q$ increases.}  \label{ccs-res2}
   \end{center}
  \end{figure}

In Figure \ref{ccs-res2}, we plot the relation between $q$ and the
number $m$ of sensors that the adversary needs to capture to
compromise a link between two non-captured nodes with probability
$p_{\textnormal{compromised}}$. We also fix the key ring size at $K
= 40$, and the key-setup probability at $p_q = 0.1$, and set
$p_{\textnormal{compromised}}$ as $0.03$ or $0.1$. The curves on how
$m$ changes with respect to $q$ also confirm the analytical results
as explained below. As mentioned, for small $m$, as $q$ increases,
$p_{\textnormal{compromised}}$ asymptotically
first decreases and then increases. Hence, if we fix a small
$p_{\textnormal{compromised}}$, as $q$ increases, the required $m$
should first increase and then decrease. This clearly is validated
by the curve of $p_{\textnormal{compromised}}=0.03$ in Figure
\ref{ccs-res2}. Similarly, the theoretical result shows that for
large $m$, as $q$ increases, $p_{\textnormal{compromised}}$ asymptotically always increases. Thus, if we fix a large
$p_{\textnormal{compromised}}$, as $q$ increases, the required $m$
should always decrease. This is validated by the curve of
$p_{\textnormal{compromised}}=0.1$ in Figure \ref{ccs-res2}.

%
%

\begin{figure}[!t]
\begin{center}
  \includegraphics[scale=0.5]{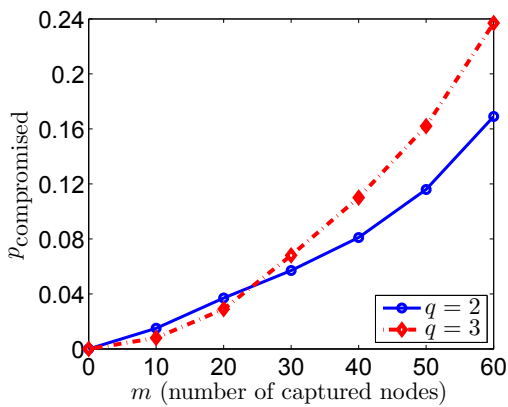}
\caption{A plot of probability $p_{\textnormal{compromised}}$ of
link compromise with respect to $m$, under $K = 50$, $P = 10000$
and $q = 2, 3$. Note that the key-setup probability here is
different for different $q$.}  \label{ccs-res3}
  \end{center}
  \end{figure}

\subsubsection{Varying $m$ with key ring size and key pool size fixed}

We now fix both the key ring size and the key pool size, and vary
$m$ (the number of nodes captured by the adversary) to see how the
probability $p_{\textnormal{compromised}}$ of link compromise
changes with respect to $m$.

Figure \ref{ccs-res3} depicts probability
$p_{\textnormal{compromised}}$ with respect to $m$. We fix the key
ring size at $K = 50$, and the key pool size at $P = 10000$, and
set $q$ (the required amount of key overlap) as $2$ or $3$. Clearly,
for each $q$, we observe that $p_{\textnormal{compromised}}$
increases as $m$ increases. In addition, as we see, for very small
$m$, probability $p_{\textnormal{compromised}}$ under $q=2$ is
larger than that under $q=3$. We explain below that this confirms
the theoretical result. By analysis, for small $m$ and small $q$, as
$q$ increases, $p_{\textnormal{compromised}}$ asymptotically
decreases. Therefore, for small $m$, $p_{\textnormal{compromised}}$
under $q=2$ is asymptotically larger than that under $q=3$, which is
in agreement with Figure \ref{ccs-res3}. By analysis, for large $m$,
as $q$ increases, $p_{\textnormal{compromised}}$ asymptotically
increases. This is validated since we see in Figure \ref{ccs-res3}
that for large $m$, $p_{\textnormal{compromised}}$ under $q=2$ is
small than that under $q=3$.

\section{Conclusion }
\label{sec:Conclusion}

In wireless sensor networks, the
$q$-composite key predistribution scheme is a widely studied
mechanism to secure communications. Despite considerable studies on secure
sensor networks under the $q$-composite scheme,
most of them ignore one or more than one of the following aspects:
node-capture attacks, sensor mobility, physical transmission constraints, the boundary effect of network fields, and link unreliability, whereas we consider all of them in this paper and present conditions to ensure secure connectivity.
Moreover,  few researches on the
$q$-composite scheme
formally analyze the scheme's resilience to arbitrary node-capture attacks considered in this paper, where the adversary can capture an arbitrary set of sensors. In the presence of arbitrary node-capture attacks, we derive  conditions
to ensure unassailability and unsplittability in secure sensor networks under the
$q$-composite scheme.
This paper presents a rigorous and comprehensive analysis to provide useful guidelines for  connectivity and resilience
design of secure sensor networks.

\end{document}